\documentclass[useAMS,usenatbib,usegraphicx]{mn2e}
\usepackage[T1]{fontenc}
\usepackage[british]{babel}
\usepackage[varg]{txfonts}
\usepackage{rotating}
\usepackage{dcolumn}

\title[Counterparts of supersoft X-ray sources in M31]{Multiwavelength search for counterparts of supersoft X-ray sources in M31}
\author[E. Chiosi et al.]
{E. Chiosi,$^{1}$ M. Orio,$^{1,2}$, F. Bernardini$^{3,4}$, M. Henze$^5$, and N. Jamialahmadi$^6$ \\
 $^1$ INAF--Osservatorio di Padova, vicolo dell' Osservatorio 5,
   I-35122 Padova, Italy \\
 $^2$ Department of Astronomy, University of Wisconsin, 475 N. Charter Str., Madison WI 53704\\
 $^3$ Department of Physics \& Astronomy,
 Wayne State University, 666 W. Hancock St., Detroit, MI 48201, USA \\
 $^4$ INAF, Osservatorio di Capodimonte, Salita Moiarello, 16, 81131 Napoli, Italy \\
 $^5$ European Space Astronomy Centre, P.O. Box 78, E-28691 Villanueva de la Canada, Madrid, Spain \\
 $^6$ Observatoire de la Cote Azur, Boulevard de l'Observatoire, 06300 Nice, France
\\
}
\begin{document}

\date{Accepted . Received; In original form }

\pagerange{\pageref{firstpage}--\pageref{lastpage}} \pubyear{}

\maketitle

\label{firstpage}

\begin{abstract}

We searched optical/UV/IR counterparts of seven supersoft X-ray sources (SSS) in M31
in the Hubble Space Telescope (HST) ``Panchromatic Hubble Andromeda
Treasury'' (PHAT) archival images
 and photometric catalog. Three of the SSS were transient, the other
four are persistent sources. The PHAT offers the opportunity to identify SSS hosting very
massive white dwarfs (WD) that may explode as type Ia supernovae in single degenerate binaries,
with magnitudes and color indexes typical of symbiotics, high mass close binaries, or
systems with optically luminous accretion disks. We find evidence that the transient SSS were
classical or recurrent novae; two likely counterparts we identified are probably 
 symbiotic binaries undergoing mass transfer at
a very high rate. There is a candidate accreting WD binary in the error circle of one of the
persistent sources, r3-8. In the spatial error circle of the best
studied SSS in M31, r2-12, no red giants or AGB stars are sufficiently luminous in the optical
and UV bands to be symbiotic systems hosting an accreting and hydrogen burning WD.
This SSS has a known modulation of the X-ray flux with a 217.7 s period, and 
we measured an upper limit on its derivative,
$\mid\dot{P}\mid\lesssim0.82 \times 10^{-11}$.
This limit can be reconciled with  the rotation period
of a WD accreting at high rate in a binary with a
few-hours orbital period. However, there is no luminous counterpart with color indexes typical
of an accretion disk irradiated by a hot central source. Adopting a semi-empirical
 relationship, the upper limit for the disk optical 
luminosity implies an upper limit of only 169 minutes for the orbital period of the WD binary.
\end{abstract}
\begin{keywords} binaries: close, galaxies: individual: M31, galaxies: stellar content, ultraviolet: stars, white dwarfs, X-rays: stars
\end{keywords}

\section{Introduction}
Supersoft X-ray sources (SSS) emit X-rays primarily at energy
below 0.8 keV and with bolometric luminosity from about 10$^{36}$ erg s$^{-1}$ to a
few times 10$^{38}$ erg s$^{-1}$, and many of them have unabsorbed luminosity 
 $> 10 ^{37}$ erg s$^{-1}$. In first approximation, the spectrum can
be fitted with a blackbody at temperature from 10$^5$ to 10$^6$ K.
 
The first SSS were discovered with Einstein \citep{Long81}, and established
as a class thanks to ROSAT \citep[see][]{Greiner00}. They are defined  phenomenologically,
and although some supernova remnants (SNR), black holes in binaries and even background 
 active galactic nuclei (AGN)
may appear as SSS, the vast majority of the SSS identified at other
wavelengths turned out to be white dwarfs (hereafter WD; see e.g.
Orio et al. 2010, Orio 2013). Very young single WD, PG1059 stars or planetary nebulae nuclei,
may be SSS for a few years. However, because the atmosphere of an
accreting and hydrogen burning WD becomes extremely hot, and
can peak in the X-rays, most SSS are WD in close binaries (hereafter
CBSS). In fact, more than half of the CBSS are post-outburst novae
with short lived SSS phase \citep[1 week to 10 years; see discussion
 of][]{Pietsch06,Orio2012}.
 CBSS comprise
also some steady sources, that have remained X-ray luminous
for most of the time in the last $\simeq$30 years.
The first observations of optically identified CBSS indicated
that the X-ray emission was indeed due to the central source. 
The X-ray gratings have revealed  WD atmospheres at
extremely high temperature, close to a million K for post-outburst
novae (see review by Orio, 2012), but also 
sources at lower effective temperature T$_{\rm eff}$,
bordering with the EUV range. Moreover,  
 in some novae in outburst and in a few persistent 
 CBSS, the stellar continuum is not observed.
 In fact, a complex emission spectrum with prominent emission lines
 in the supersoft X-ray range may mimic a stellar
 continuum, (e.g: MR Vel, Bearda et al. 2002, Motch
et al. 2002, or CAL 87, see Greiner et al. 2004, Orio
 et al. 2004), but  in broad band spectra of CCD-type detectors the emission lines are not resolved 
(see Orio 2012). This type of X-ray spectrum has been
 attributed to mass outflows or to the nova wind
 shocking circumstellar material   \citep[see e.g.][]{Ness05,Bearda02}.
 The central source temperature in some of these cases has been estimated to be 
 in the range T$_{\rm eff}$=30,000-150,000 K,
 but it may be difficult to determine, depending much on intrinsic
 absorption.
   \begin{figure}
   \centering
   \resizebox{\hsize}{!}{\includegraphics{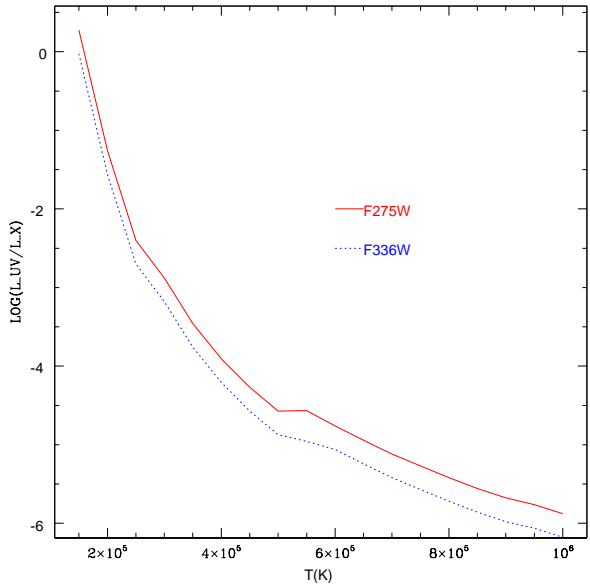}}
 \caption{Ratio of WD flux in the F275W filter UV range and  flux in the 0.2-1 keV X-ray range
 (red) and ratio of WD ``U'' flux in the F 336W filter and  flux in the 0.2-1 keV X-ray
range
 (blue) for a given effective temperature of the WD atmosphere, here approximated with a blackbody.}
   \label{ratios}
    \end{figure}

We know that most SSS are post-outburst novae, or other interacting
white dwarf (WD) binaries; much less often SSS have
turned out to be SNR or
AGN (see Orio et al. 2010). In order to cast light on the nature
of the sources we need to identify their counterparts at other wavelengths.
The majority of the Galactic or Magellanic Clouds SSS, including
 those that have not been observed in a nova eruption and seem
 to be permanent sources, are
binary systems with a WD accreting hydrogen rich material from
a companion and burning hydrogen in a thin shell. These system
may be type Ia Supernovae (SNe Ia) progenitors and have attracted
attention for this reason. Novae and SNe Ia undergo similar
evolution. In both cases the WD accretes material on the surface,
but in the first case the hydrogen accumulated on the surface
reaches fusion into helium resulting in envelope ejection and rise
in optical magnitude by at least 8 mag, without disintegration of
the WD, while in the second case the Chandrasekhar mass limit is
reached, causing a global explosion. It is unlikely that the majority
of novae end as SNe Ia, although a few recurrent novae (RN; novae
with repeated outbursts during a century) are promising SNe Ia
progenitors' candidates due to their high WD mass.

M31 is the most massive galaxy in the Local Group and has a
distance modulus $\Delta$m-M=24.45 (Dalcanton et al. 2012). It has been
extensively observed in X-rays with ROSAT, Chandra and XMM-Newton,
and it hosts a conspicuous population of SSS. The number
of observed SSS was already of  the order of a hundred in 2009 (see Orio et al.
2010, and references therein), while for comparison 
 in the much closer Magellanic Clouds, only 29 SSS have ever been 
 detected. A little more than a half of the SSS
in Andromeda have turned out to be post-outburst novae, and two
of them are SNR (Orio et al. 2010, and references therein). The
nature of many others is still unknown. In this paper we 
search optical/UV/IR counterparts for seven sources whose
fields have already been observed in the PHAT (Dalcanton et al.
2012). Only one of them has a possible nova counterpart, while
others have never been identified with novae. The coordinates of
the sources, the HST exposures and epochs of X-ray detections or
non-detections are given in Table 1.

 The rest of the paper is organized as follows:
 in Section 2 we discuss the PHAT, 
 the positional and photometric errors of its catalog, and the potential to observe
 SSS in its deep images. We also explain which specific types
 of CBSS we can expect to observe. In Section 3 we present the characteristics of the
 SSS with optical counterparts in the PHAT and analyse the short term variability
 of the most luminous and better observed of these sources, r2-12.
 In Section 4 we present the photometric results, obtained by matching
 the public PHAT catalog in the different filters. The catalogs differ slightly in
 positions and the fields are very crowded, so we had to accurately match the sources 
 observed in the different filters.
In Section 5 we discuss the possible counterparts or lack of thereof, and Section 6 contains our conclusions.

   \begin{figure}
\includegraphics[width=85mm]{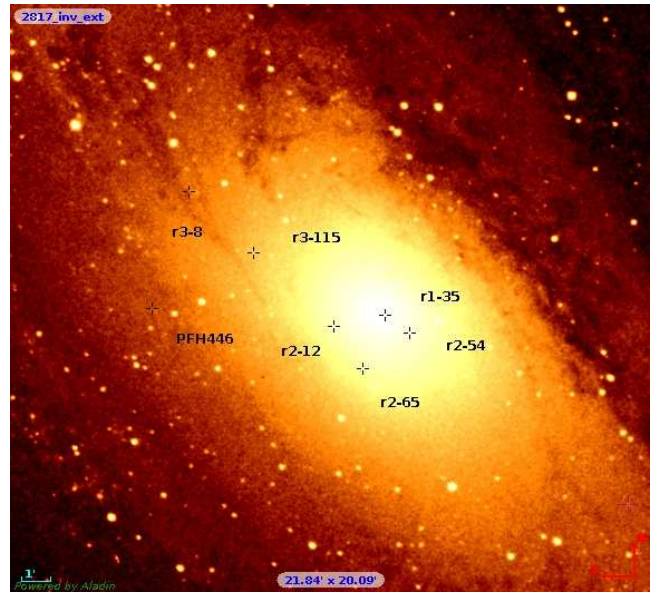}
    \caption{Positions of the SSS examined
 in this article in M31 on an Asiago photographic
 plate taken by the late L. Rosino (East is on the left and North on top). 
 The center of the Figure is at equatorial
 coordinates $\alpha_{ 2000}$=00,43,06.58 and $\delta_{2000}$=41,13,46.98. 
}
   \label{map_m31.fig}
    \end{figure}

\section{The PHAT, a precious resource}
The HST archival data we used are the PHAT
images and catalog (Dalcanton et al. 2012). The PHAT is an HST
multi-cycle Survey of Andromeda, at present in its third year, that
at completion will have imaged 1/3rd of the M31 star forming disk,
covering a contiguous one square degree area in 828 orbits, with 6
different filters: 2 optical ones (F475W and F814W) used with the
ACS/WFC camera, 2 ultraviolet ones (F275W and F336W) used
with the WFC3/UVIS camera, and 2 infrared filters (F110W and
F160W) used with the WFC3/IR camera.

The survey area is divided in several so called ``Bricks'' which in turn contain different
``Fields''.
 As Table 1 shows, most SSS counterparts we searched
 are located in Brick 1 that covers the central 
 (and obviously very crowded) region of M31. 
 Source r3-8 is located instead in Brick 7.
We used the VEGA magnitudes of the
 photometric catalog published by the PHAT collaboration
 (http://archive.stsci.edu/pub/hlsp/phat/).
 The likely optical counterpart of one of the SSS is also detected in the Local Group Survey
 \citep[LGS, see][]{massey2006}.

The astrometric accuracy gives an absolute error of
 100 mas (dominated by errors in the ground based reference catalog) and an error of 10 mas
 in the relative positions of objects detected in all the three cameras.
 
The photometric measurements are obtained with  a
50\% completeness at signal to noise of 4 at F275W=25.1,
 F336W=24.9, F475W=27.9,
 F814W=27.1, F110W=25.5, F160W=24.6 for single pointings in the uncrowded outer disk.
 In the inner disk the optical and the NIR data are crowding limited and the deepest
 reliable magnitudes are up to 5 mag brighter \citep{dalcanton2012}.
In this paper we will not make use of the F160W filter, for which the
 photometric catalog errors are large and crowding is severe in the central
 region of M31 in which our SSS are located. 
The photometric median errors for  given measured magnitudes are given 
 by Dalcanton et al. (2012) in their Fig. 13, upper left panel,
 for Brick 1 in which 6 of the sources are located. Source r3-8 is instead in
 Brick 7, which has similar errors to Brick 9 (upper right panel of Fig.
 13 in the Dalcanton et al.  paper). Basically,
 while the measurements are crowding-limited in the optical and IR filters, 
 in the U and UV UVIS filter the measurements are limited by photon counting.
 Therefore, the UVIS photometric accuracy is not strongly dependent on crowding.
 The median error
 is always larger than 0.01 mag in the F275W filter, but it is still smaller than 0.02 mag up
 to Vega magnitude $\approx$23.8 in the F475W filter.
 However, the median errors are large above approximately 24th magnitude; specifically,
above Vega magnitude 25, the median error is larger than 0.2 mag in the F275W filter, 
 it is close to 0.2 mag in the F336W and F814W filters, and it is about 0.1 mag in the
 F475W filter, for which the best precision is obtained. 

\subsection{CBSS detectable in the PHAT} 
In M31 we have to rely mainly on photometry
 to identify possible counterparts of X-ray sources,
because most optical counterparts are  
 too faint for ground based optical spectroscopy with the current
 facilities.

  The PHAT is of high interest for the study of SSS in M31, because the upper limits 
for detection above the 4$\sigma$ level, reported above, are sufficiently
 deep to reveal interacting binaries accreting through a disk (or even via a wind
in the case of symbiotics),  with  
 $\dot m \geq 10^{-8}$ M$_\odot$ year$^{-1}$
 necessary for steady nuclear burning.
The effective temperature of
 an accreting and hydrogen burning WD varies in a large
 range depending on WD mass, mass transfer
 rate $\dot m$, and  on whether the source is observed after
 a thermonuclear flash (Starrfield et al. 2012). The WD 
 itself would be detectable above the PHAT upper limits
 for the F275W and F366W filters without further contribution
 of an accretion disk or a symbiotic nebula only if it is on
 the low side of  effective temperature distribution, with
 T$_{\rm eff}\leq$200,000 K. This can clearly be seen in from Fig. 1, where 
 assuming the blackbody approximation for the atmosphere, we show
how the ratio of X-ray flux to U/UV flux dramatically
 changes with the blackbody temperature.
 Above about 200,000 K the WD is so hot, that the Raleigh Jeans tail of the
 flux distribution moves to the far and then to the extreme UV.

 In fact, all the Magellanic Clouds SSS known to be CBSS with 
a hydrogen burning WD have visual magnitude M$_{\rm V}\leq$1.3, which
 translates into V=25.75 in M31, or slightly fainter due to reddening 
 \citep[see measurements listed by ][]{orio1994,simon2003}.
 The brightest non-symbiotic known CBSS in both X-rays and optical is CAL 83,
 although at optical wavelengths RX J0513.9-6809 is of comparable magnitude and
 sometimes more luminous. For CAL 83 MACHO and OGLE light curves
 have been published \citep{Greiner2002,Raj2013}.
 Assuming a distance modulus to
 the LMC $\Delta(m)$=18.47 \citep{Freedman2012} and A$_{\rm V}$=0.249 \citep{burstein82},
 we find that the V magnitude varies between V=-0.92 and V=-2.02, 
 due to a a 0.4 mag orbital magnitude amplitude superimposed
 on additional, discrete steps  in average magnitude
 in three different ``states'' (high, intermediate and low). 
 Simultaneous multi-color light curves in all the states have not been measured,
 but low states \citet{simon2003} reports (U-B)$_0 \simeq$-1.17 and
 (B-V)$_0 \simeq -0.05$. We do know that at least the V and R magnitudes decrease
 or increase by the same amount in the different states \citep{Greiner2002}.
 Clearly, a source like CAL 83 would easily be picked up in the PHAT.

 {\v S}imon (2003) tried to to characterize the
 color indexes of all the SSS known in the Galaxy and the Local
 Group as a class.  
 Not all the objects included by {\v S}imon are of
 interest as examples for our search,
 because the author included in his Table 1 objects he calls ``VSS'', 
 which have never been observed in X-rays as SSS
 and may be luminous only in the far UV. 
 V751 Cyg, a VY Scl star, is defined as a SSS only because it showed
 a ``softish'' spectrum in an observation done
with the HRI \citep{Greiner2000b}, but the spectral resolution
 was very poor and the source was underluminous for an SSS. We know now that 
 variables of this class generally are not SSS in the low state  \citep{Zemko2013}.
 The symbiotics listed by {\v S}imon were not observed
 in X-rays, although they were selected because they show indications of a very hot central
 source. The general indication we draw from
 this paper are that the majority of known SSS in the Galaxy and
 in the Magellanic Clouds would be observable in the PHAT
 if they were at M31 distance, although some measurements
 may have large errors (see Fig. 3). 

 A very interesting results of the study of SSS in the Local Group in recent years has
 been the discovery of SSS in high mass X-ray binaries containing a WD
\citep[see review by][and references therein]{orio2013}.
 Novae eruptions are observed in binaries with  low mass companions,
 but no known
 evolutionary effect works against the formation of stable accretion
in high mass
 binaries with a WD, which would
 have outbursts of smaller amplitude. We know now that hydrogen burning in high mass WD binaries
 must be instead relatively
 common, since four SSS in the Magellanic Clouds   have been identified
 with massive binaries.
 Of these systems, only MAXI J0158-744 in the SMC was observed to undergo an optical
 brightening at the time of the X-ray flare \citep{li2012}.
 For a significant number of SSS in M31 the only possible counterparts detected so
 far are hot O and B stars \citep[see][]{orio2010}, suggesting that there a large class
 of SSS host a WD and a young, hot massive companion (thus, they
 are High Mass X-ray Binaries, or HMXB).

 The other subclass of SSS that should really easily stand out in the PHAT
 are the symbiotics, with their large luminosity. 
 In the optical and IR we detect all AGB and red giant stars in the PHAT.
 In order to select criteria to identify possible symbiotic counterparts,  
 we report the available
 photometric measurements for symbiotics that have been observed as SSS
 in the first four columns of Table 2. We later compare them with possible
 counterparts that are U/UV luminous, also included in the Table.
 Most interesting are those symbiotics which we
 believe host a steady burning WD:  they are not only
 blue objects, but they are also ``red'', hosting a red giant, supergiant or AGB secondary.
The colors and the optical and UV luminosity of the three known SSS-symbiotics  
 stand out, because they are very different than those of most other  stars.
 The large luminosity in the B and U filter is due to a bright
 accretion disk or at times only to a wind from which the WD accretes,
 because the donor star must supply the hydrogen rich
 material to the WD at a rate at or above
 10$^{-8}$ M$_\odot$ year$^{-1}$. These symbiotics
 appear at least as hot and blue as A type
 stars, but usually have lower luminosity than  A-type stars in the V filter.

 Concerning the transient SSS, a classical nova that has returned to quiescence
 would remain undetectable in the PHAT, but a recurrent nova (RN) in a symbiotic
 system would still be observable at quiescence. The difficult problem is how to identify it
 only on the basis of color indexes. In the Galaxy more than half of the observed RN are in
 symbiotics of the RS Oph type. 
 What do these objects look like after  hydrogen burning has ceased?
 If they  accrete through a disk, the accretion disk is no longer irradiated,
 and the surrounding nebula is not photoionized anymore by an extremely hot central
 source.   ``Normal'' symbiotics that are not known to be undergoing
 hydrogen burning, or where
 the WD is obscured by heavy intrinsic absorption,
 are still luminous objects  in many wavelengths.  Their colors
 are different from the ones of the luminous cool component alone,
 but there is a large spread in the range of
 magnitudes and color indexes.
 They are luminous in the red filters, 
but also in the F475W and  U/UV filters, because of the hot component. 
 In the K band, symbiotics have absolute magnitude in the range
 -2.5 and -7.5 \citep{munari2002}.
  The expected  UV magnitude range in the F275W filter can be inferred from
 the UV magnitude at the near peak
wavelength (2600 \AA) indicated by \citet{kenyon1984} for
 Galactic symbiotics: It is about 1-1.5 mag fainter than the V magnitude.
 The magnitudes in the Kenyon \& Webbink article were not corrected for reddening,
  and since the average absorption of the Galactic symbiotics 
 is higher than towards M31, we expect even a smaller difference
 between optical and UV magnitudes in Andromeda.
The known Magellanic Clouds symbiotics are unusually bright, apparently
 containing mostly AGB stars, in the range V=(-2.2)--(-3.4) and
 with B-V$\leq$1.3.  All 
symbiotics are very luminous in H$\alpha$ \citep[see][and discussion therein]{goncalves2006}.
 We obtained H$\alpha$ images of the M31 core with the WIYN telescope (see e.g.
 Orio et al. 2010) and also examined the archival 
  H$\alpha$ images in the LGS,  but we found that the transient
SSS of our group of sources are in field effected by excessive diffuse light
for ground based photometry.

 \begin{figure*}
   \centering
\parbox{7.5cm}{
\resizebox{7.5cm}{!}{\includegraphics{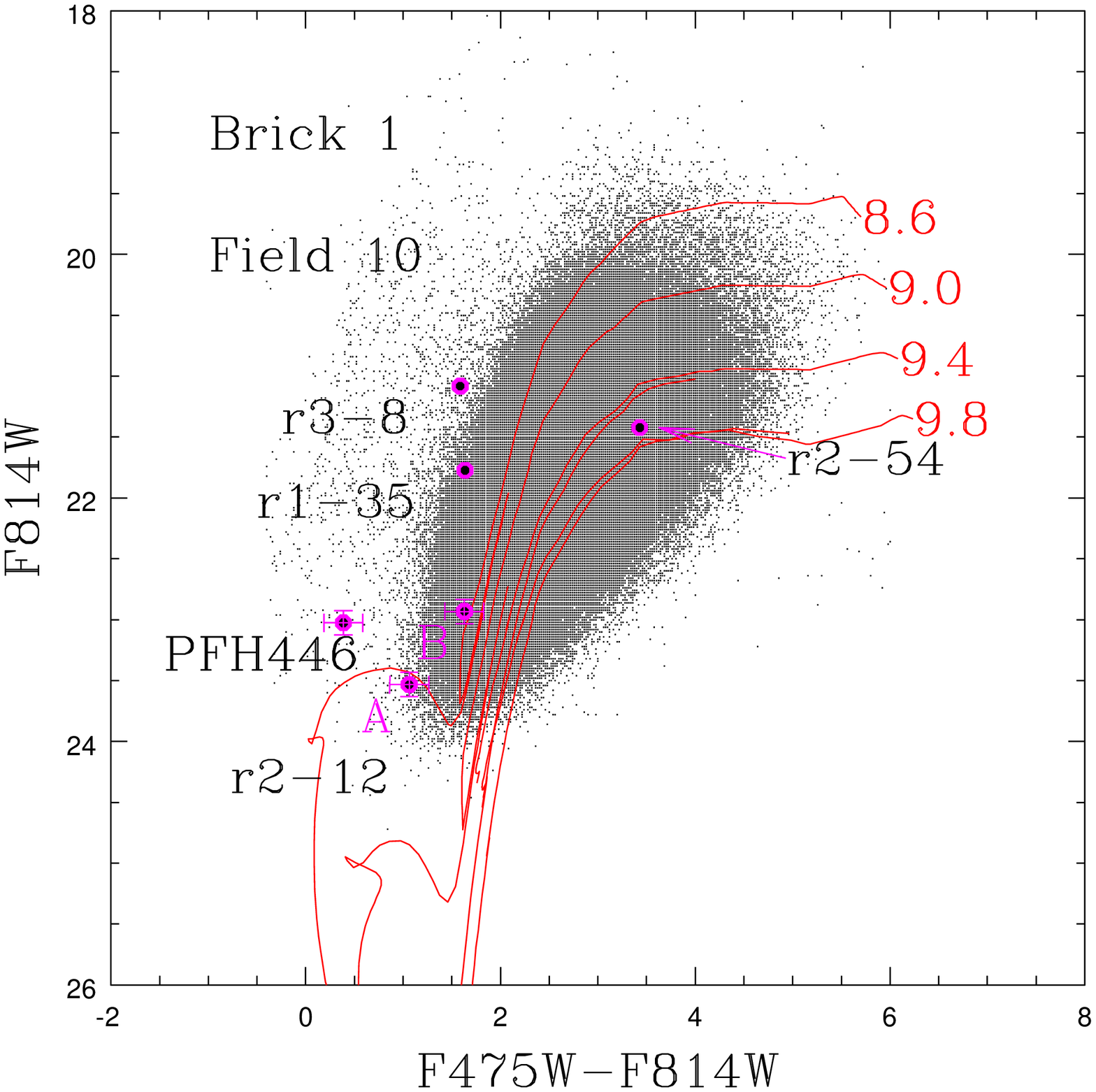}}\\
}
\hspace{0.2cm}
\parbox{7.5cm}{
\resizebox{7.5cm}{!}{\includegraphics{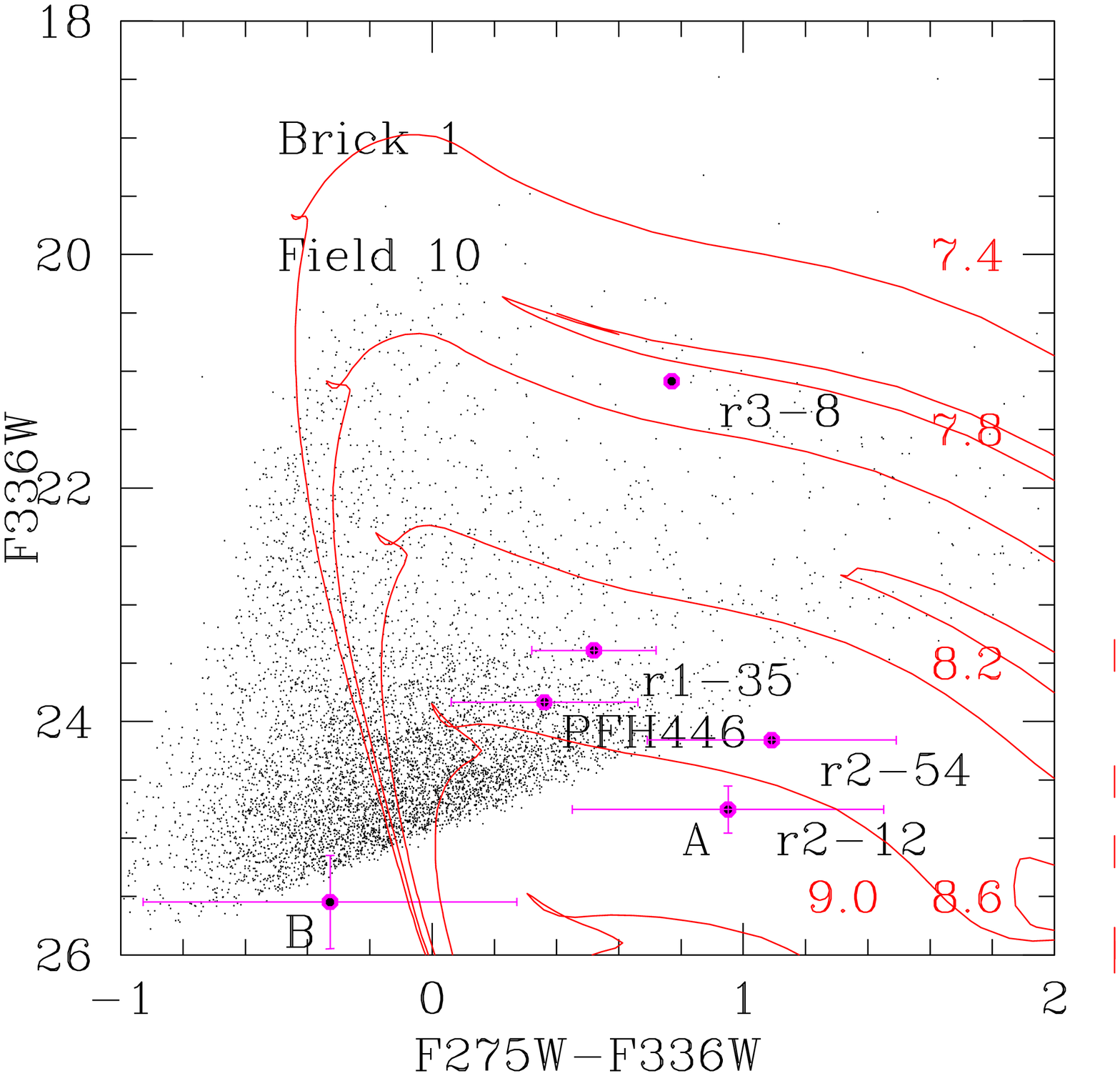}}
}
    \caption{A set of  isochrones ranging in age from  9.8 Gyr
  to to 7.4 Myr superimposed on the color
 magnitude diagrams in the optical and UV
 filters of the field 10 of Brick 1, containing r2-12. We also show the positions of the
 ``candidates'' discussed below. We have ``reddened'' the isochrones
 assuming E(B-V)=0.08. We have cut the optical color magnitude diagram to include only
 the reliable measurements, without the objects in the
 larger photometric catalog with less than 4 $\sigma$ detections.
 In the
 UV, because the
 measurements are not crowding
 limited, we have included
 the measurements with large error bars.
}
   \label{iso.fig}
    \end{figure*}
\section{A sample of intriguing SSS}
 Fig. 2 shows the position of the SSS we investigated, on the scan of a plate of
 M31 taken with the Asiago 1.8m telescope by L. Rosino. 
 All objects are
 within 5 arcmin of the inner core of M31. At the time 
 we wrote this article, there were sufficiently deep
 HST archival images to observe the optical counterparts of seven SSS in M31,
 four of which are persistent sources (with some 
 short lasting, recurring ``off'' states for r3-8). 
 The persistent sources, likely to be steady accretors, are 
 are most likely detectable at other wavelengths. 
 Since all optical novae undergo an SSS phase (albeit at times
 effected by excessive intrinsic absorption for detection) and they are 
 the largest class of SSS in M31 \citep{pietsch2005,orio2010,henze2013}, we must be aware that
 most transient SSS do belong to this class, and their optical
 counterpart may not be recovered in the PHAT in many cases. 

 The color magnitude diagrams (CMD) of the objects
 in Field 10 of Brick 1 of the PHAT (containing
 the source r2-12), obtained from the catalog of Dalcanton et al. (2012),
is shown in Fig. 3.
 This field is quite central, crowding
 is particularly severe, so the upper limit magnitudes above a
 4$\sigma$ level in optical
 and IR are quite higher than in the outer regions: about 
 magnitude 25.1 in the F475W filter,  magnitude 24 in the F814W filter,
 and only around  magnitude 20 in the F110W filter. 
 From the theoretical isochrones also plotted in Fig. 3 we can immediately see that,
 despite previous conclusions that there
 is a mixture of populations in the M31 inner core,
 young stars in this extended inner core region are very
 rare and almost all the population has been formed between 8.6 and 9.4 Gyears ago.
A complete discussion of this result can be found in \citet{dalcanton2012}. 
 High mass binaries are thus unexpected in this region, and 
 indeed none of the SSS 
 we are going to describe turned out to be a member
 of this interesting subclass.

In the following context we use the names of the {\sl Chandra} catalog adopted
 among other authors by \citet{distefano2004}. Source
PFH 446 was observed only with {\sl XMM-Newton}, and we refer to the catalog of the 
 paper by \citet{pietsch2005}. 
The X-ray detections and some characteristics of these SSS have been reported by
 \citet{distefano2004,orio2006, orio2010}.

\subsection{SSS observed in the PHAT}

 Our SSS can be divided in three groups: persistent hot and luminous sources,
persistent sources at the low end of the luminosity distribution,
 and transient objects that, as we already mentioned, are likely to have
 been optical novae.
 
The first of the persistent ``dim'' sources, 
{\bf r2-54} \citep{distefano2004}, was only marginally detected in
 an observation of 2001 and many later observations \citep{hofmann2013}. It was
 still detected at 
 at the beginning of 2011, although \citet{hofmann2013} find that it should 
 lower absolute luminosity
 than the canonical SSS (around 10$^{35}$ erg s$^{-1}$) and several of the {\sl Chandra}
observations between 2006 and 2011 yielded only upper limits  \citep[see][]{hofmann2013}.
The source confusion in the {\sl XMM-Newton} fields
 prevents clear detection, so we rely only on Chandra data for this source.

{\bf r2-65},
initially detected in 2001 \citep{distefano2004, orio2006, orio2010}, 
 is also very close to the M31 inner core.
In Orio (2006) and  Orio et al. (2010) it was described as a transient source. However,
Hofmann et al. (2013) show that it was often observed again in {\sl Chandra} HRC-I images
between 2008 and 2012, and  with the latter authors' assumptions on N(H) 
 it would also have X-ray luminosity below 10$^{36}$ erg s$^{-1}$, like r2-54
\citep{hofmann2013}. The non-detections have upper limits above 10$^{36}$ erg s$^{-1}$,
 so the luminosity may have decreased below this threshold when it was
 not detected.

The first of the transients, {\bf r1-35},
 was first noticed as a possible candidate remnant 
 of the nineteenth century  supernova S And of 1885 as discussed by 
 \citet{distefano2004, hofmann2013}. The supernova
 identification had already been examined and
rejected by \citet{kaaret2002}, who noted instead a transient bright counterpart
 in HST images of June 1995, a likely nova in outburst. 
\citet{pietsch2005} suggested the association
    with Nova M31 1995-09b, which is in the same position of Kaaret's transient.
Given that M31 can be observed only at the end of the night in the Summer
 and that many ground based telescopes in the Northern hemisphere are effected
 by poor weather in the Summer, it is very likely that the nova was discovered
 from the ground only in September although the outburst occurred in June or earlier. 
 The supersoft transient was observed only almost 3 years later, but
 this is not unusual, especially for the M31 novae \citep{henze2013}.
The X-ray luminosity of r1-35 was only
 $\simeq 3 \times 10^{35}$ erg s$^{-1}$ \citep{hofmann2013},
 which may be  consistent with a nova caught during the decline from peak X-ray luminosity,
or with a soft
 X-rays emission lines spectrum from the ejecta \citep[e.g. T Pyx,][]{Toffle2013}.
It is also consistent with the overall M31 nova population
statistics presented in Henze et al. (2013).

{\bf r3-115} was a peculiar transient SSS: it appeared as a typical very soft SSS in
{\sl Chandra} ACIS-S observations in 2001, but 107 days later it showed also
 a hard component. The X-ray spectrum became harder when it decreased in luminosity,
 more consistently with a black-hole transient than with a nova or other
WD binary. A possible association
with Nova M31 1998-07d seem unlikely because of the spectral evolution and the time lag
(if a hard component is observed, it is usually in a fast nova, e.g. RS Oph).

{\bf PFH 446} was a transient only detected in {\sl XMM-Newton} observations
\citep{pietsch2005}  and the spatial uncertainty
in the position is much larger than for the {\sl Chandra} sources,
 about 2 arcsec.
  
Finally, there are two very luminous and better studied sources, r2-12 and r3-8.
{\bf r2-12}
 is a very luminous, persistent SSS in the central region of M31. 
 It has two important characteristics:
 a modulation with a period of 217.7 in the X-ray light curve, 
 \citep{trudolyubov2008, orio2010},
and, assuming it is a WD, a  very hot atmospheric temperature, $\simeq$
 700,000-900,000 K in different
 exposures (Orio et al. 2010 and references therein, Henze et al. 2013).
 The luminosity varies from a few times 10$^{37}$ erg s$^{-1}$ to a few
 times 10$^{38}$ erg s$^{-1}$.
 The supersoft X-ray flux modulation   
 is likely to be the spin period of a massive WD, spun up by accretion.
 Another possibility is that of non-radial g-mode pulsations 
\citep[see e.g. discussion by][]{leibowitz2006}. These short period oscillations
 are frequent in SSS where all evidence points at  ``canonical'' hydrogen burning WD. 
 Multiple periods  were observed in the supersoft X-ray source
 post-nova V4743 Sgr; they were attributed to both WD rotation and non radial oscillations
(Leibowitz et al. 2006). The RN RS Oph and KT Eri 
 in the supersoft X-ray phase displayed a modulation with a $\simeq$35 s period
 \citep{beardmore2010,osborne2011}, a period of 54 s has recently been measured in nova V339 Del
 \citep{beardmore2013, ness2013},  and the X-ray SSS flux of CAL 83 is modulated with a 67 s period \citep{odendaal2013}. 
 It is very intriguing that r2-12 has been X-ray
luminous for about 25 years, since the
 first observations  of M31 with {\sl Einstein},
 and it has always been observed as one of the most
 luminous X-ray sources in all pointings of the M31 core with all imaging X-ray telescopes,
 albeit with a modest level of variability \citep[see e.g.][]{hofmann2013}. 
 In the next Section we briefly analyze the variation of the X-ray
 period in the X-ray database of HEASARC, and discuss a limit  on
 the period derivative.

The only other long lasting SSS with  near-Eddington luminosity and
 effective temperature T$_{\rm eff}$
 above 400,000 K are CAL 83 in the LMC
and the symbiotic SMC 3 in the SMC. CAL 83 seems to be a
 massive WD with a companion with M=1.5-2 M$_\odot$
\citep{vandenheuvel1992},  an orbital period of a day \citep{smale1988} and
T$_{\rm}\simeq$550,000 K \citep{lanz2005}. The X-ray flux of
 CAL 83 is modulated with  
 a 35 min period \citep{schmidtke2006} in addition to the  
 67 s period mentioned above.
 Occasionally, this source shows an ``X-ray off'' state for at least a week.
 SMC 3 has an orbital period
 of 4.5 years and T$_{\rm eff} \simeq$450,000 K (Orio et al.
 2008 and references therein). It also shows brief
 states of very low X-ray luminosity, always at the same orbital phase.
 Both sources, like r2-12, were initially 
observed with {\sl Einstein} and were always observed as SSS throughout 
 the last $\simeq$30 years. 
 Despite some variability, r2-12 has always appeared above the detection
 threshold of the existing observations.

{\bf r3-8} is a luminous variable source: it usually has an X-ray luminosity
 above 10$^{37}$ erg s$^{-1}$, but it also displays ``off'' or ``low''
states, falling below luminosity detection thresholds (depending on the image) of
 10$^{34}$-10$^{35}$ erg s$^{-1}$  for 
a few weeks at a time \citep{orio2010}.  It has been detected 
 as one of the bright M31 sources in
 a very large number of observations
since {\sl ROSAT} times. The association with an 
 H$\alpha$ emitting object of the LGS \citep{massey2006} has been
 suggested by \citet{hofmann2013}. This 
 proposed counterpart is in front of a dust lane, where background sources have an extremely 
 low probability of detection \citep[see][]{orio2010}. 

\subsection{The 217.7 s modulation in the X-ray flux of r2-12}
Given the rich database of X-ray observations in the HEASARC archive,
 stretching through more than 25 years and including
 frequent exposures in the last 10 years, we
 measured the X-ray period in a subset of X-ray exposures with the aim of determining
 a period derivative and constrain the models. Trudolyubov \& Priedhorsky
 follow a model proposed by \citet{king2002} to explain a longer period 
 SSS observed in M31, described in detail by \citet{osborne2001}. The latter, however
   turned out to have a very brief transient
 life (it was not detected again after a few months)
and therefore it was almost certainly
a nova \citep[see discussion by][]{orio2010}.  
Trudolyubov \& Priedhorsky assume that r2-12 is accreting as an intermediate polar (IP), residing in 
 a binary in which the donor is more massive of the WD, as in the model
 of \citet{vandenheuvel1992}. In the King et al. scenario,
 accretion is occurring through a Keplerian disk, but the WD has a sufficiently
 strong magnetic field to be on its way to becoming a polar, where eventually
 accretion occurs only to the polar caps. Even in this disk-fed stage, the authors
foresee that 
the polar caps are significantly hotter and more luminous than the rest of
 the WD atmosphere (even if thermonuclear burning may be occurring everywhere at the
 base of the envelope accreted by the WD), so the modulation
 of the flux is due to temperature difference at the poles. Assuming
 $\dot m = 5 \times 10^{-7}$ M$_\odot$ year$^{-1}$,
 an accretion rate that 
 explains the supersoft X-ray luminosity as due to nuclear burning
 fueled by high $\dot m$, the maximum spin up rate to
 transfer all the accreting material momentum to the WD is  
     $\mid\dot{P}\mid\lesssim0.065$ s yr$^{-1}$.  

The periodic signal is not detectable in all the observations
 we examined. 
Four {\sl XMM-Newton} observations spaced by 1 day and lasting 
from 20 to 30 ksec were done in July
 of 2004, but the period is detected above a 3 $\sigma$ significance level only on 2004 July 16
 (see Table 3).  
  In order to constrain the measurement
 of the periodic signal, and to assess if it changed
in time, we examined the available pointings carried out with {\sl XMM-Newton} and with
 the  {\sl Chandra} HRC-I camera, choosing those 
with an exposure time longer than $\sim20$ ks, 
(shorter exposures result in estimates affected by high uncertainty).
We did not examine any 
 {\sl Chandra} ACIS-S images, where the source measurements 
 yield a low count rate; neither the ACIS-I images
 in which the count rate is even much lower for this soft source.
We extracted the two longest HRC-I exposures
 a pointing done on 10-31-2001 (47 ksec) and
 one of the 2011 July observations
(21 ksec) with CIAO V4.6, but we 
did not measure any period with a significance above 3 $\sigma$  
 in these images.
Thus, at this stage, we decided to examine 
{\sl XMM-Newton} exposures, in which
 the count rate is higher. The details are reported in Table 3.

 We avoided pointings in which solar flare contamination was dominating
in more than the half of the exposure time. We obtained a baseline
of approximately 12 years with roughly one pointing per year, with a total of 10  observations.
 The {\sl XMM-Newton} data were processed with XMM-SAS version 13.0.0, 
 with the latest calibration file (CCF) available in November 2013.
During all observations, both the pn and the MOS cameras were operating in full frame mode,
 but the analysis reported here is only based on the pn, which yielded a much
 higher S/N and has a time resolution of 73.4 ms compared to 2.6 s of the MOS.
Standard data screening criteria were applied in the extraction of scientific products.
We excluded
 from the analysis the intervals of solar flares, and used the screening
 criteria \textsc{PATTERN=0} and \textsc{FLAG=0} to obtain light curves in the 0.15-1 keV
 range (the source has practically no residual flux at higher energy).
We extracted the source photons from a circular region of radius 20" centered at the best source
position, with the exception of the few cases in which the source is extremely close to the 
edge of the CCD,
where instead we used a circular region of radius 12.5".
The background was selected from a circular region of 40" in a region in the same CCD 
as the source lies.
The source event arrival time of each observation, in the 0.15-1 keV energy range 
 were converted into barycentric dynamical times
with the SAS tool \textsc{barycen}.

 We first computed the power spectrum and verified the presence of a
peak around $4.6\times10^{-3}$ Hz,  then we estimated the periodicity of
the peak by means of a phase-fitting technique \citep[see][for details]{dallosso2003}.
 Due to the sparse
observational coverage, the measurements, shown in Table 3, are independent and not phase connected.
 We note that the results for the first four observations in Table 3
 are fully consistent with those of \citet{trudolyubov2008}. 

Fig. 4 shows the evolution of the period with time.
 We first performed a fit with a constant
period, obtaining a period of $217.76\pm0.05$ s with a 
 reduced  $\chi^{2}_{\nu}=0.57$ for 9 degrees of freedom.
 We concluded that
the signal was constant within the uncertainty.
 The  3 $\sigma$ upper limit on the derivative of the period,  
obtained by performing a fit with a constant plus a linear function,
is $<8.2\times10^{-10}$.
This upper limit corresponds to $\mid\dot{P}\mid\lesssim0.026$ s yr$^{-1}$, almost a factor of 2.5
 lower than the estimate obtained assuming the the Trudolyubov \& Priedhorsky model. 
 In the model assumed by these authors, an IP with an orbital period of several hours, 
 the upper limit on the period derivative  translates into an upper limit in the mass accretion rate, 
$\dot m \leq$ 10$^{-7}$ M$_\odot$ year$^{-1}$, which is consistent with 
 steady hydrogen burning \citep[e.g.][and references therein]{fujimoto1982, vandenheuvel1992}.
\begin{figure}
\begin{center}
\includegraphics[angle=-90,width=8.5cm]{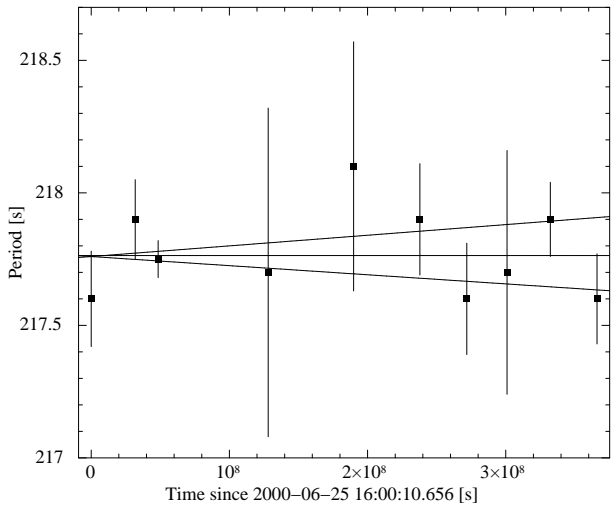}
\caption{X-ray period of r2-12 as a function of time.
The central solid line shows the fit with a constant value $217.75\pm0.05$ s,
 while the the top and bottom lines represent
the 3$\sigma$ limits on the period derivative $\mid\dot{P}\mid<=8.2\times10^{-10}$.
 The reference time corresponds to the mid-exposure time 
of the first analysed dataset (observations no. 0112570401 of  {\sl XMM-Newton}).}
\label{fig:period}
\end{center}
\end{figure}
    
\section{The photometric results}
Using the PHAT photometric catalog, in Fig. 4, 5 and 6 we plotted two 
  sets of CMD, the  optical ones
 (already seen for r2-12) and the UV, for Brick 1 in which most
 of our SSS are located. The plots also show the measurements for
 some candidate counterparts that we will examine further below.  
 The CMD diagrams on the left are almost equivalent to 
 the often used CMD plots of I vs. (B-I) and
 B vs. (U-B), since the used HST filters have almost 
 the same bandpass as these Johnson filters. However, caution is needed in the comparison because
 translating the system of the HST filters to the Johnson system is 
 rather complex and a precise transformation formula has not been found. For details
 see \citet{Sirianni2005}.
The angular dimensions
 of each optical field are 202''x202'', for the WFC3+UVIS
 the dimension is 162''x162'', and for the WFC3 IR channel
 it is  123''x136''. 
 In Fig.  4, 5 and 6 we indicate the sources in
 the spatial error circle of each SSS with the red circles.
We also plotted the position of SMC 3, which has the same reddening
 as most of these SSS (see Table 2), for comparison.
We adopted Cardelli's et al. extinction law
\citep{Card1988} and an average  value E(B-V)$\sim$ 0.08 mag
 in Brick 1, while we assumed E(B-V)$\sim$ 0.10 for source PFH 447 in Brick 7.
These values are based on work done by \citet{unwin80} and although
we expect that more precise estimates will become available in the next years
 \citep[see e.g., for the central 2200 pc,][]{Dong14}, the 
 identification of possible counterparts is not affected very significantly  
 by the value of the reddening we adopt, which is always small. 
 Comparisons with the best candidates for five of the SSS are shown in Table 2.

 Table 4 (available on line) shows the multi-band photometric
 measurements for the objects detected in HST images within the spatial error circle
 of the SSS.
 We immediately note the absence of optical counterparts more luminous than 19th mag
in all fields,
 and more luminous than 21st mag in several of them. Moreover, the vast
 majority of the possible counterparts are red stars,
 with colours typical of Asymptotic Giant Branch (AGB), red giant and supergiant
 stars in M31. We will discuss the
 exceptions below.

 Generally, a foreground X-ray source emitting very soft X-ray flux would be more luminous than 21st magnitude
 at optical wavelengths. The first class of very soft foreground X-ray sources are 
 AM Her systems, or polars. In a survey
 of ROSAT-identified magnetic CV by \citet{Pret2013}, all polars within 750 pc
 from the Galactic plane are at most 10 times
 more X-ray luminous than r2-12. Polars emitting the same X-ray flux 
 as r2-12, but located in the Galaxy (thus with orders
 of magnitude lower absolute X-ray luminosity than r2-12), 
 have an optically bright counterpart at least in the F474W filter. 
  In the error circle of r2-12 the only
 optical counterparts measured with sufficient
 precision have colors of red giants (or subgiants, as we discuss below). 
 We conclude that   
the lower limits on L$_{\rm X}$/L$_{\rm opt}$ ratio for all SSS here examined,
including the two dim ones, 
 are larger than any foreground polars (see also discussion by King et al. 2002
for another M31 SSS ). 

 The other
 class of very X-ray soft foreground sources are cooling neutron stars, with their thermal
 emission. However, they  are observed only within $\simeq$300 pc from the
 Galactic plane, and at such distance also they 
 would have  detectable counterparts with much ``bluer'' colours than the red giants. 
 Perhaps the less optically bright neutron stars with a thermal component may be
 Geminga, which was discovered because of his gamma-ray emission and pulsars' properties,
 but according to some authors has a second component originating on the neutron star surface
 \citep{karga05}.
 At about 250 pc, Geminga has an optical counterpart with R$\simeq$25.5, still
 within the PHAT detection limits. 
 Thus, a first
 conclusion is that foreground sources are very unlikely counterparts for the whole group of SSS
we are examining here.

\subsection {Astrometry}
For most objects we report the positions of \citet{kaaret2002},  who
improved the 
absolute positions of sources r2-12, r3-8, r2-65, r1-35 and r3-115 
in the Chandra images of the High Resolution Camera (HRC, uncertainty of 1'')
 registering  them with objects in the 2MASS catalog, so that the final coordinates
 differ from the 2MASS positions by at most 0.4''. We 
 estimate that this is approximately a 1$\sigma$ positional uncertainty, although
 Kaaret further improved by applying an average
 coordinate shift between the HRC and 2MASS of 0.2'', obtaining an average
 systematic
 displacement of only 0.15'' from the 2MASS positions. Kaaret's positions differ by up to 0.23'' by
 the positions determined by Hofmann et al. (2013) in other,
 more recent HRC-I images for the same sources. The 1$\sigma$  uncertainty
 estimate in the positions of Hoffman et al. (2013) is 0.38''.
 The position of two of the persistent sources, r2-12 and r2-54, 
were also measured in the Chandra ACIS camera observations of \citet{kong2002}
 and we adopt the ACIS position for r2-54. These 
 authors registered the images using
 the absolute positions determined by Kaaret (2002), thereby also obtaining a maximum
 difference from the 2MASS positions of 0.4''. We note that
 the r2-12 position measured by  Kaaret 
 is  0.33'' distant from the  one obtained by Kong et al., although
 it is within the error circle of the former.

The release notes of the HST Legacy archive indicate 
 an error of about 0.4" in the HST/2MASS  image registration. By  combining 
 two 1 $\sigma$ errors of 0.4", we 
evaluate that our 1 $\sigma$ error circle has a radius of $\simeq$0.6" in the 
relative Chandra/HST positions. For Brick 9 we found the largest
 difference from the 2MASS positions (a systematic shift of about 0.08" in right
 ascension and 0.2" in declination in the U/UV images) and we re-registered the images
 with MASS.  For PFH 446, the SSS that was only
 observed with {\sl XMM-Newton}, the 
 source the position uncertainty is about 3''.
 All the  1 $\sigma$ relative position uncertainties
 estimates we gave here are very conservative.

\section{The candidate counterparts}

 In Table 4, which is available in the online Journal, we list the objects detected 
 in the spatial error circles of 6 of the SSS of our sample, in regions    
 with photometric measurements published by the PHAT collaboration
 (Dalcanton et al. 2012). The ``hottest'' object, with the larger (U-B) and/or
 (UV-U) color index, has been listed first, while the other objects are
 in order of increasing declination. 

\subsection {The persistent SSS}

 In Table 2 we report UV/optical/IR absolute magnitudes,
 and color indexes for three well studied symbiotics of the
 Local Group that are steady SSS, and  
 for several candidate counterparts of our SSS. We have chosen the 
 apparently ``hotter'' object in the fields. However, 
 candidate ``A'' in the r2-12 error circle.
is {\it not} a good candidate SSS, although it is the the
 ``bluest'' optical object within 0.7 arcseconds of the {\sl Chandra}
 position of r2-12 detected with the UV filters.   
 All the other objects in the spatial error circle of r2-12 have typical 
 colors of evolved cold stars, except the ``B'' candidate
 of Table 4 and Fig. 2, which however
 was measured with large errors in the U filters
 and does not appear to be sufficiently
 luminous in the F275W filter to be an SSS.

 As expected, young hot stars are absent. 
Table 2 also shows the 
 (B-J)$_0$ color index, useful for comparison with supergiants of OB type
 (see Wegner et al. 1993).
 If r2-12 was a symbiotic system, it would have a counterpart with color indexes 
 close to those of SMC 3 
 (see discussion above and Orio et al. 2007). Lin 358, Draco C-1,
 and AG Dra have effective temperatures around 200,000 K, which is too 
 low for a meaningful comparison with this SSS. ``Candidate A'' in
 the Table has too large a negative value of (U-B)$_0$ compared with 
 all the SSS-symbiotics in the Table, even the low temperature ones.
 For candidate ``B'' this color index is even much more negative.

 No PHAT-detected object, including
 candidates ``A'' and ``B'', stand out in an H$\alpha$ image of the field, taken with HST
for another program.
 From a comparison with the off-band, the 
 magnitude of all objects detected with this filter is in fact consistent with only continuum
 emission. All in all, in the spatial error circle of r2-12 we cannot identify the 
 ``signature'' of a hot symbiotic that may have ongoing hydrogen burning. 
Candidate ``A'' seems to be a subgiant star at $\simeq$8.6 Gyears.
 The subgiant phase is very short lived, but clearly it is not impossible to observe. Candidate
 ``B'' has been measured with great uncertainty with the F275 W filter,
 in which it appears faint.

If r2-12 is not a symbiotic, it is likely to be instead
a short period binary (P$_{\rm orb} \leq$ 1 day) with characteristics
 similar to  CAL 83, specifically high effective temperature,
 near-Eddington X-ray luminosity and persistence over many years.  
We reported that  the range of absolute measured magnitudes 
 is V=-0.92-2.02 \citep{Greiner2002}: this large optical luminosity is attributed 
 to an accretion disk,
 which is irradiated by the very hot primary. The hotter and more luminous is the WD, the more
 luminous is the disk at optical wavelength.
 Models of disks for WD binaries 
 and their luminosities are discussed in  \citet{frank1992,popham96}. 
 The optical luminosity of the known
 accreting and hydrogen burning WD binaries turns out to be proportional to their orbital
 period \citep{simon2003, vanteeseling1997}. 

 The X-ray luminosity of r2-12 at M31 distance, in the 0.2-10 keV
 range, does not fall below the value 3 $\times 10^{37}$ erg s$^{-1}$ indicated  
by van Teeseling et al. (1997) for another LMC source, RX J0439.8-6809.
This SSS in the LMC does have an unambiguous optical counterpart of magnitude V$\simeq$21.6,
 which would be undetectable at M31 distance.
The dereddened 4 $\sigma$ above background 
upper limit for the r2-12 region of Brick 1 is  M$_{\rm B}\simeq$25,1 
 and this implies an absolute
 visual magnitude  upper limit M$_{\rm F475W} \geq$+0.70
if  (B-V)$_0 \simeq$-0.05 like in CAL 83.  
Assuming there is correlation between absolute visual
 magnitude of an irradiated disk and orbital period at a given X-ray luminosity
of the type $\Sigma$=L$_{\rm X}$ P$_{\rm orb}^{2/3}$=f(M$_{\rm V}$),as 
derived by Van Paradijs \& McClintock (1994), van Teeseling et al.
(1997) analyzed
 the period and visual magnitude values for several
 SSS in the Magellanic Clouds.  They obtained a semi-empirical relationship
 M$_{\rm V}$=0.83($\pm0.25)-3.46(\pm$0.56)log$\Sigma$  
(see their Figure 3). With this relationship, the upper limit on
 the PHAT magnitude upper limit translates in an orbital period shorter than 169 minutes. 
 If we search for objects detected at less than the 4 $\sigma$ level, the
 error in the U/UV filters is larger than 0.3 mag and it is very difficult to draw any
 conclusion, so we assume that what we obtained here is the lowest limit on the
 systems we can detect.  

 r2-12 may even be an AM CVn type of object (see e.g. Bildsten
 et al. 2013 and references therein). It is tempting to attribute the 
 217 s period to a nonradial oscillation of the WD (Bildsten
 et al. 2013), but the accretor is too hot to be in the known nonradial pulsations
 instability strip. On the other hand, non radial g-mode pulsations, of a ``putative''  but still
 unexplored instability strip at much higher temperature have been attributed
 to several post-novae (see discussion by {\citet{Orio2012}).    
 
\citet{vanteeseling1997} also discuss  the constraints on the accretion disk inclination 
 that can be derived from the absence of an X-ray eclipse.
 For a source like r2-12, which is sufficiently X-ray luminous that it observed to vary, 
 an X-ray eclipse should be measurable, however it was never
 observed in long continuous X-ray observations with {\sl XMM-Newton}
 \citep{orio2010}.   
 This additional constraint indicates an inclination lower than $\leq$75$^{\rm o}$,
 assuming of course that the modulation with the 217.73 s period is due to the WD
 spin and is not the orbital period of an extremely compact binary 
 with unusually small size, such that its 
 optical luminosity would be too low for a detection. If 
 the 217 s period is due to the WD rotation as we have assumed so far,  we do not observe 
 any other modulation with the orbital period  with
 a significance above 3 $\sigma$. We thus conclude that 
 orbital modulations are not observable, so the inclination  is 
 much lower than 75$^o$ inferred from the eclipse absence
(with some assumptions, \citet{vanteeseling1997} infer 
 an upper limit of 10$^o$ from a missing orbital modulation for the SSS they
 analyzed). 

 In short, once we have ruled out all the red giants,
 supergiants and the one likely ``subgiant'' (candidate ``A''),
as likely counterparts, a missing optical
 detection, coupled with the lack of orbital modulation indicating 
 low inclination, sets strict upper limits for the optical luminosity
 of the r2-12 counterpart, ruling out many binary configurations.
 If the 217.7 s period is due the to WD spin, in Section 3 we have also shown
 that the model of an IP in a binary with an orbital period of several hours,
 proposed by Trudolyubov \& Priedhorsky, implies a higher rotational period derivative 
 than the one we measured. This may be interpreted as additional evidence that r2-12 
is in a binary with an orbital period below the period gap.

The other  
luminous persistent source in our sample is r3-8. \citet{orio2010} did
 not found any optical counterpart in a deep ground based image, but
 \citet{hofmann2013} noted the detection of an H$\alpha$ luminous object in
 the LGS. In Table 2 we indicate the values for the LGS and PHAT measurements.
 The PHAT position is about 0.5'' from the Chandra one. However,
 there is a systematic shift in the astrometry of Brick 7. 
 This candidate counterpart is also very
 luminous in U and UV. Its colors are not very different from those for SMC 3,
 fully 
 consistent with a symbiotic star with a very hot component.
Most important, the photometric measurement of the LGS and the PHAT indicate
 that the object has brightened in all filters, by 0.5 mag or more. 
 It is more luminous in the PHAT, even if,
as we show in Fig.8, the HST images indicate that the LGS measurements
 may have been contaminated by unresolved nearby objects, 
 especially in the F814W filter. In a WIYN image of 2005 Orio et al. (2010) did
not detect it below B$\simeq$23 at a 3$\sigma$ confidence
 level. It thus seem that this object is as variable at
 optical wavelengths as it is in X-rays. 

 In the crowded fields of r2-54, a source at a lower luminosity level
 than the canonical SSS, (a few
 times 10$^{35}$ erg s$^{-1}$) we found an anomalous object that is luminous in
 the red colors, but its ``UV age'' appears younger than the age inferred
 from the optical colors. We included it in Table 2 for a comparison
 with the symbiotics and symbiotic candidates.
 
The colors of all objects detected in the r2-65 spatial error circle are
only consistent with red giants, supergiants and AGB stars.

\subsection{The transients}

 As we expected, given the generally old population
 of this central region in M31, we do not find any hot young stars 
 in the error circles of our transients sources. Thus, we ruled out
 transients of the new class
 discovered by e.g. \citet{li2012} and discussed  
 by \citet{orio2013}.

If any of the persistent sources was caught during   X-ray off states,
 and the effective temperature
 decreased while the radius increased, the SSS should have become  luminous UV sources
 \citep[see][]{starrfield2012} with 
 bolometric luminosity above
 10$^{36}$ erg s$^{-1}$, resulting approximately in 
 magnitude $\leq$-2.5 in the F275W filter while
 at optical wavelengths they may have been fainter than the PHAT limits.
 There is no object with such high UV 
luminosity in any of the positional error circles of the SSS
 we examined.
 
None of the optical counterparts
 in the error circle of   r3-115 has magnitudes and color indexes
that may clearly be attributed to symbiotics, accreting WD with a luminous disk, 
 or a luminous secondary stars, suggesting that these X-ray transients were 
 novae. Most likely, accretion at the
 rate needed to sustain hydrogen burning is no longer going on.

We do find luminous red stars which are also unusually
 luminous in the blue/UV filters in
 the error circles of r1-35 and PFH 446. We detect no optical/UV counterpart
 for the optical Nova of 1995 which was associated
 with r1-35. There is an interesting object
 is located at 0.34'' from the nova position, which however is known with higher accuracy.
 The object detected in the PHAT, included in Fig. 2 and Table 2, 
 matches the color indexes of the hydrogen burning symbiotics; it is
 at only 0.3'' from the {\sl Chandra} position.
 While we think that the 1995 nova
 counterpart is much more likely to be the right one, we include this object 
 in the Table, as an interesting candidate hydrogen burning symbiotic. 

 The putative symbiotic in the error circle of PFH 446, like the symbiotic
 candidate for r2-35 has 
 absolute IR magnitude $\simeq$-2 in the F110W filter, 
 consistent with a red giant 
 (see Fig. 4-6 and Table 2). However, the magnitude in the the F475 W filter and in the UV
ones is larger than expected, probably indicating a second component
producing the optical and UV flux.
The counterpart in the error circle of PFH 446 is
 at only  0.4'' from the {\sl XMM-Newton} position, at
 2000 coordinates $\alpha$=00,43,25.54, $\delta$=+41,16,17.4.  
As the plots in Fig. 7 show, this object is on the blue side of 
 the red giant branch. For the
dereddened absolute magnitudes and color indexes we assumed E(B-V)=0.1
 and Cardelli et al.'s dependence of extinction on wavelength \citep{Card1988}. 
 A search in the LGS H$\alpha$ images 
 made us rule out strong  
 H$\alpha$ emission above the continuum level. 
 Symbiotics do have strong
  H$\alpha$ emission; thus, the jury is still out on the nature of this candidate
 counterpart.

\begin{figure*}
   \centering
\parbox{7.5cm}{
\resizebox{7.5cm}{!}{\includegraphics{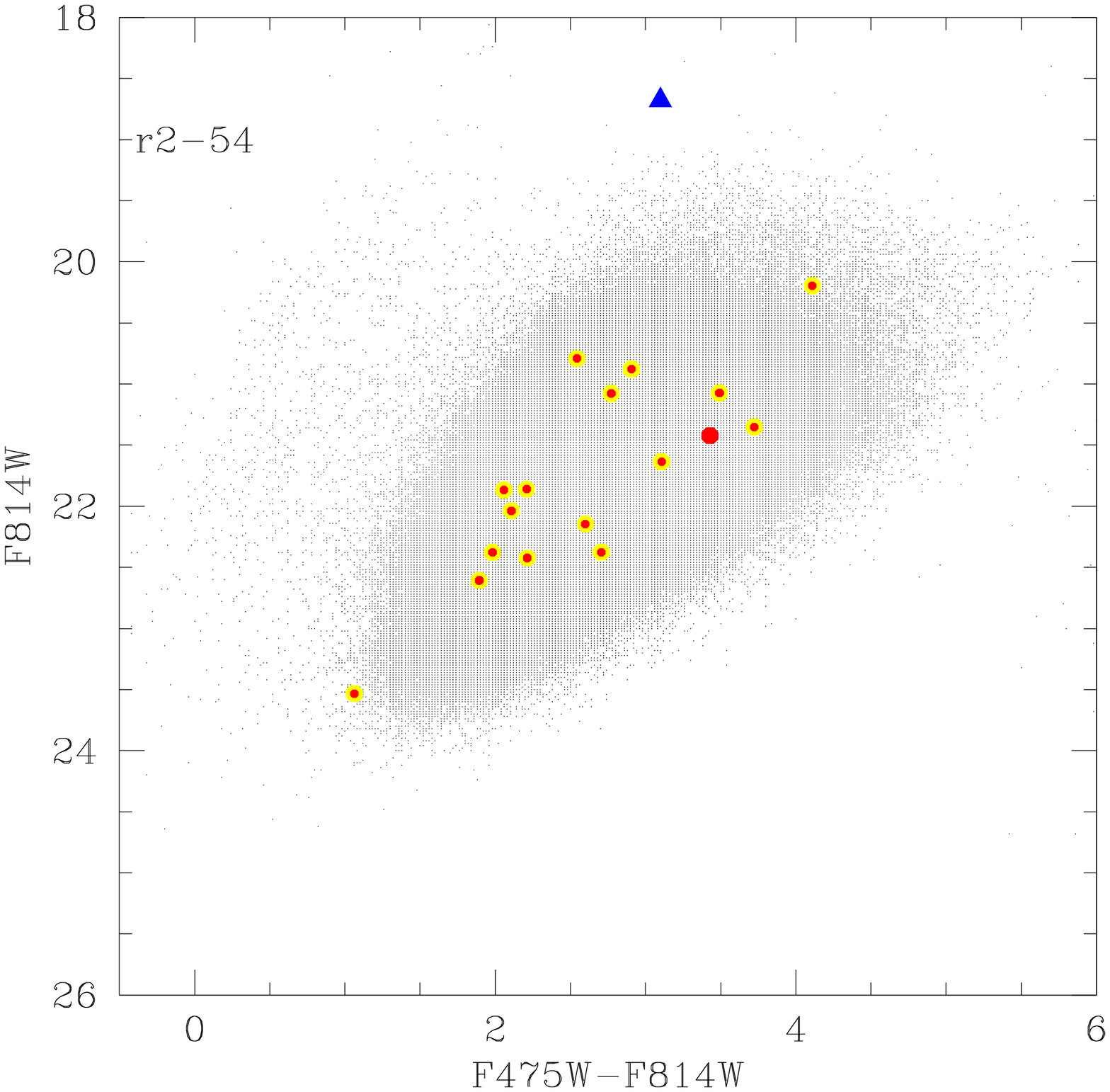}}\\
}
\hspace{0.2cm}
\parbox{7.5cm}{
\resizebox{7.5cm}{!}{\includegraphics{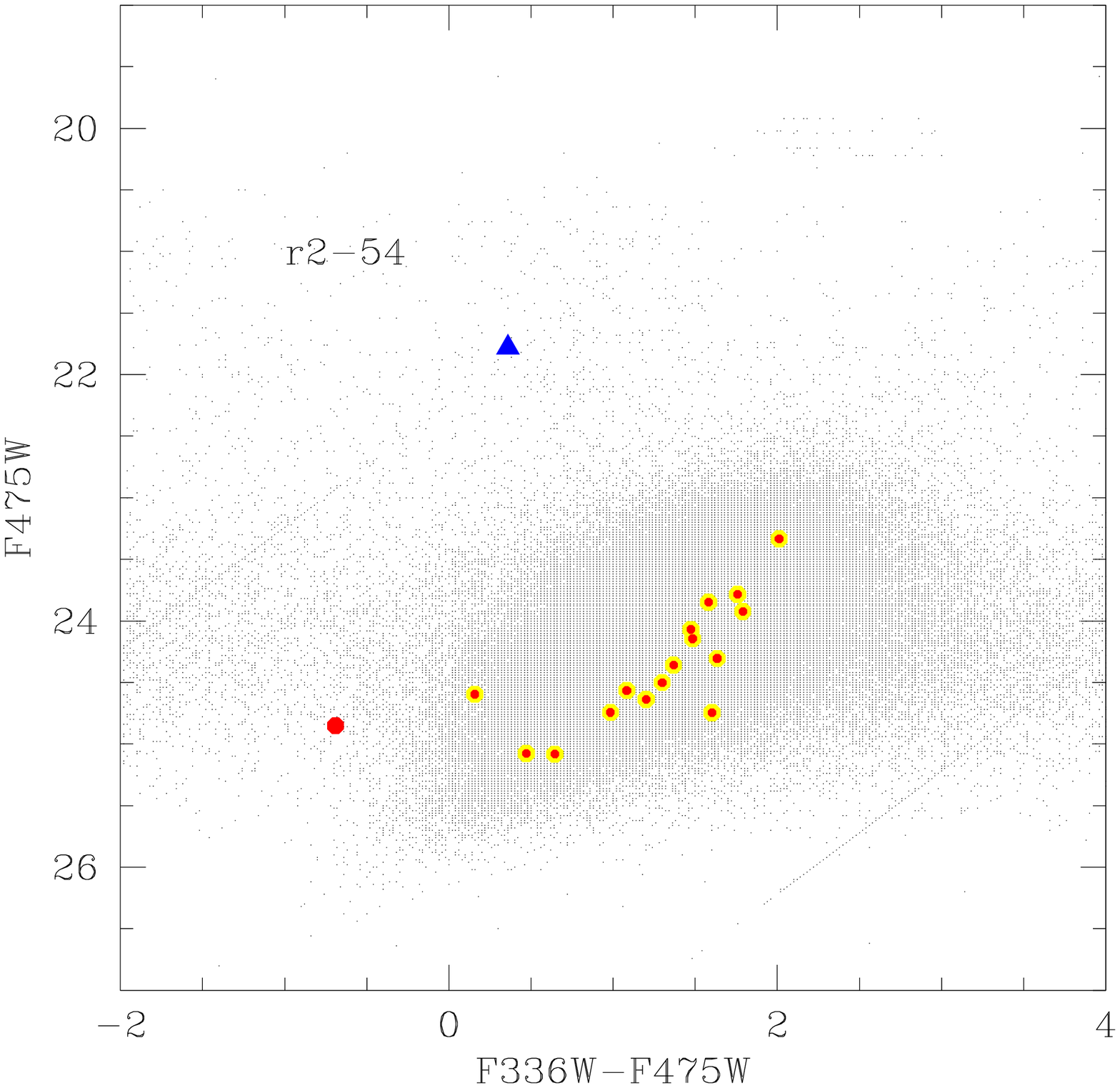}}\\
}
\parbox{7.5cm}{
\resizebox{7.5cm}{!}{\includegraphics{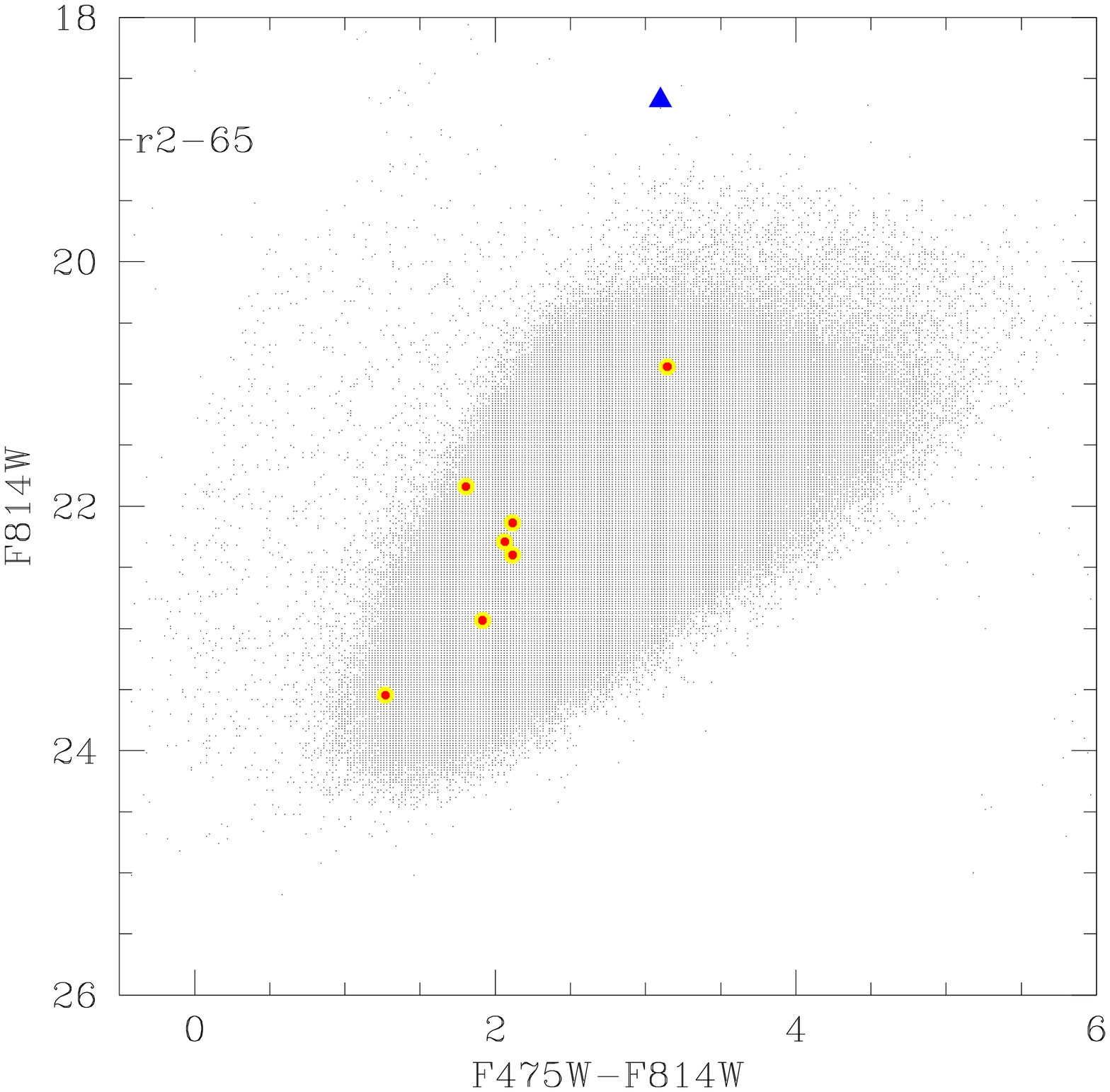}}\\
}
\hspace{0.2cm}
\parbox{7.5cm}{
\resizebox{7.5cm}{!}{\includegraphics{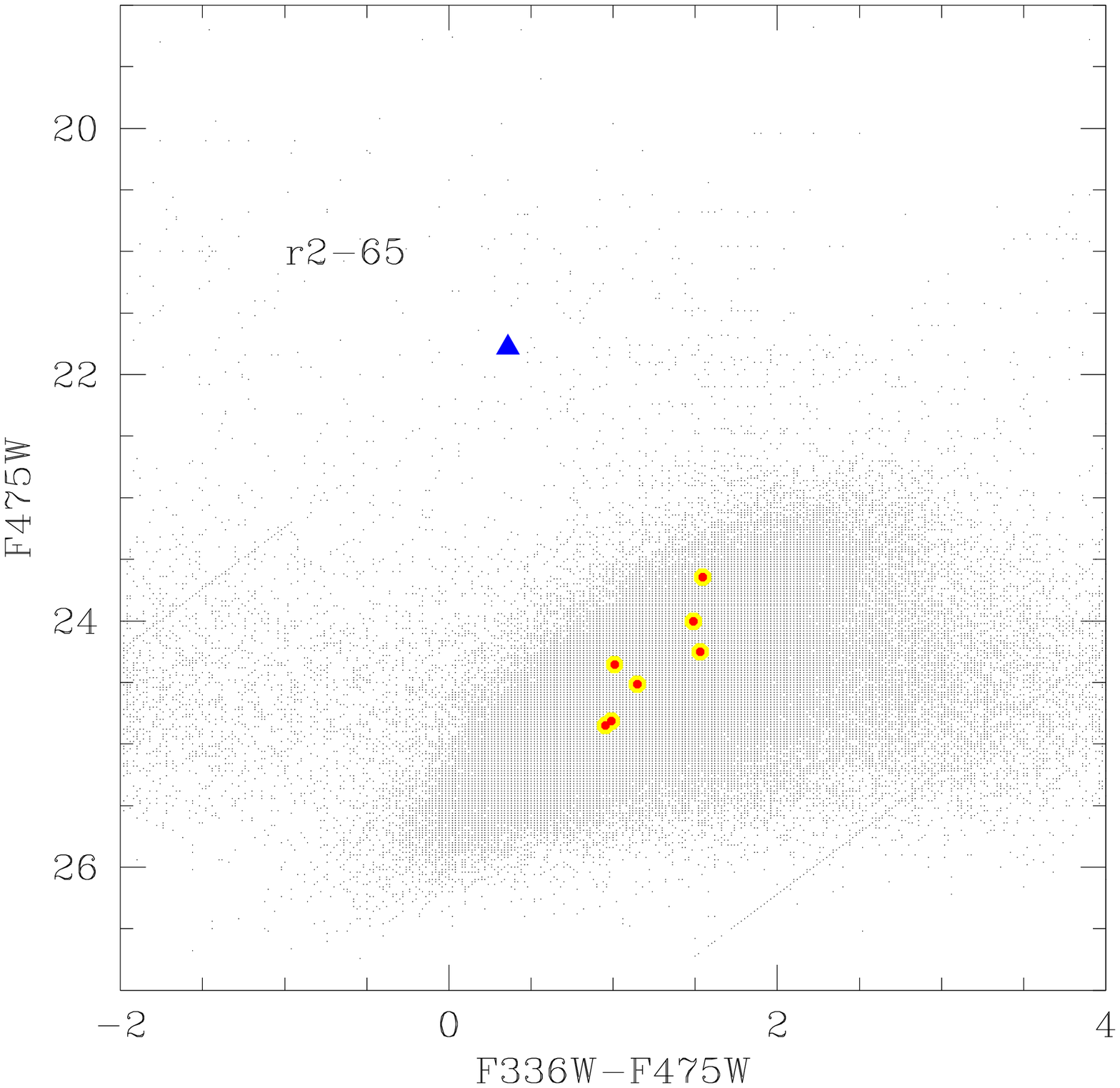}}\\
}
 \caption{The position of the optical/UV counterparts in the error circle of the 
``dim'' persistent SSS on the optical and U/UV CMD of the whole observed field, 
based on the Dalcanton et al. photometry. The blue triangle represents the position on the CMD of the symbiotic star SMC3  as it would appear at M31 distance. We are mainly looking for luminous red stars placed to 
 the left than the majority of the other objects of comparable luminosity, in this plot and 
 in Fig. 5 and 6. The most likely candidate for r2-54 is indicated 
 with a red dot.}
\label{cmd1.fig}
\end{figure*}
\begin{figure*}
   \centering
\parbox{7.5cm}{
\resizebox{7.5cm}{!}{\includegraphics{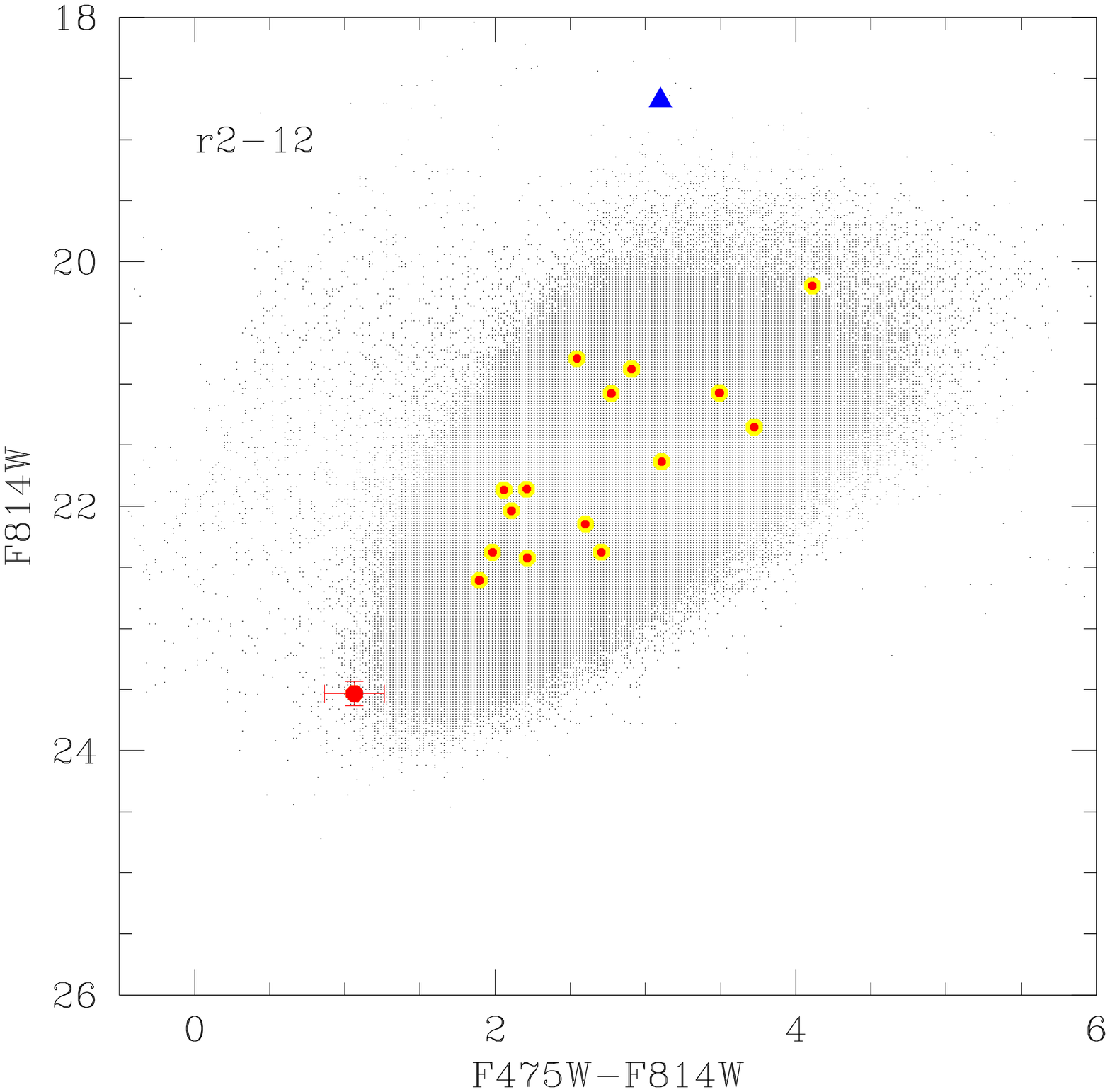}}\\
}
\hspace{0.2cm}
\parbox{7.5cm}{
\resizebox{7.5cm}{!}{\includegraphics{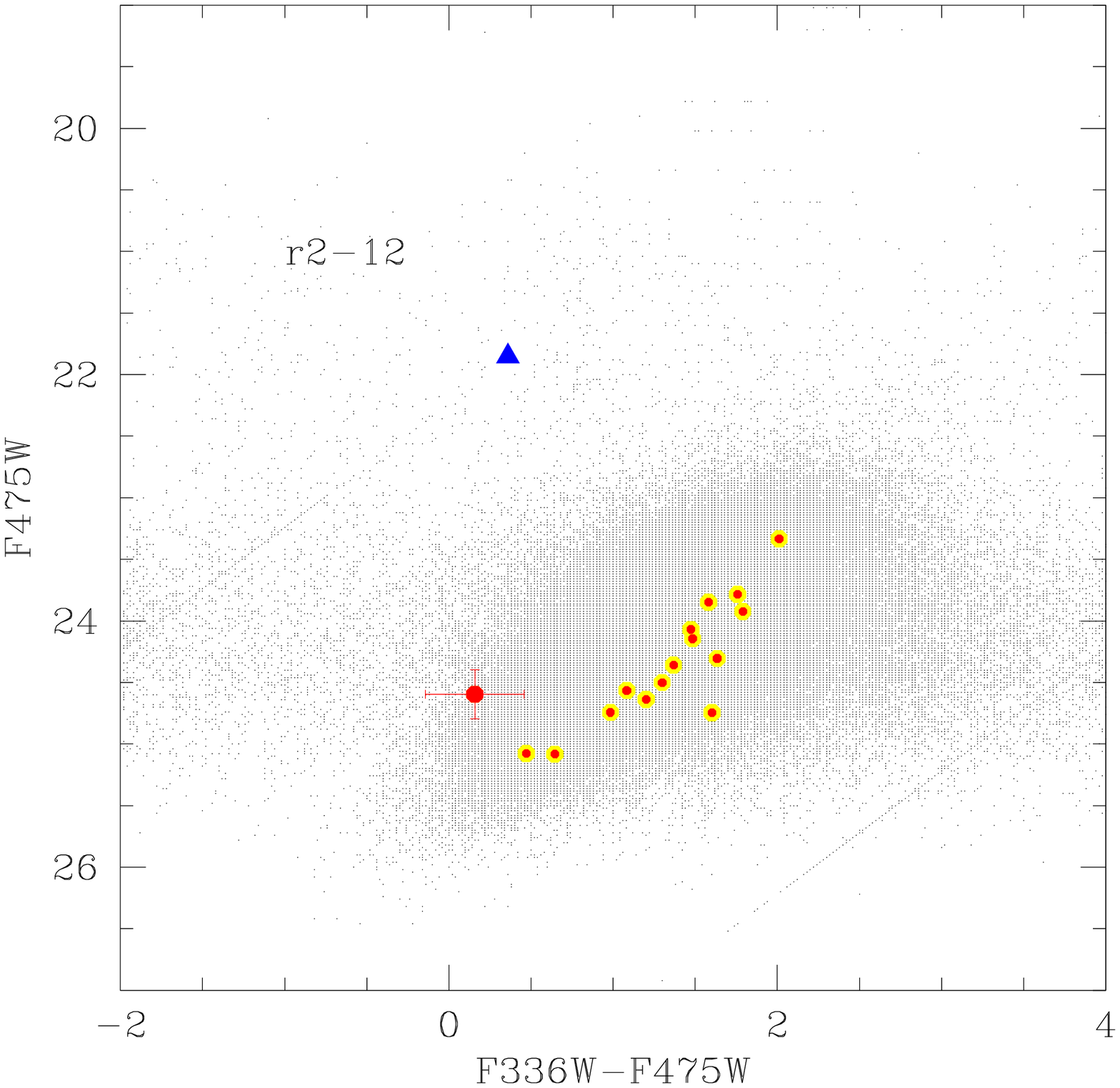}}\\
}
\parbox{7.5cm}{
\resizebox{7.5cm}{!}{\includegraphics{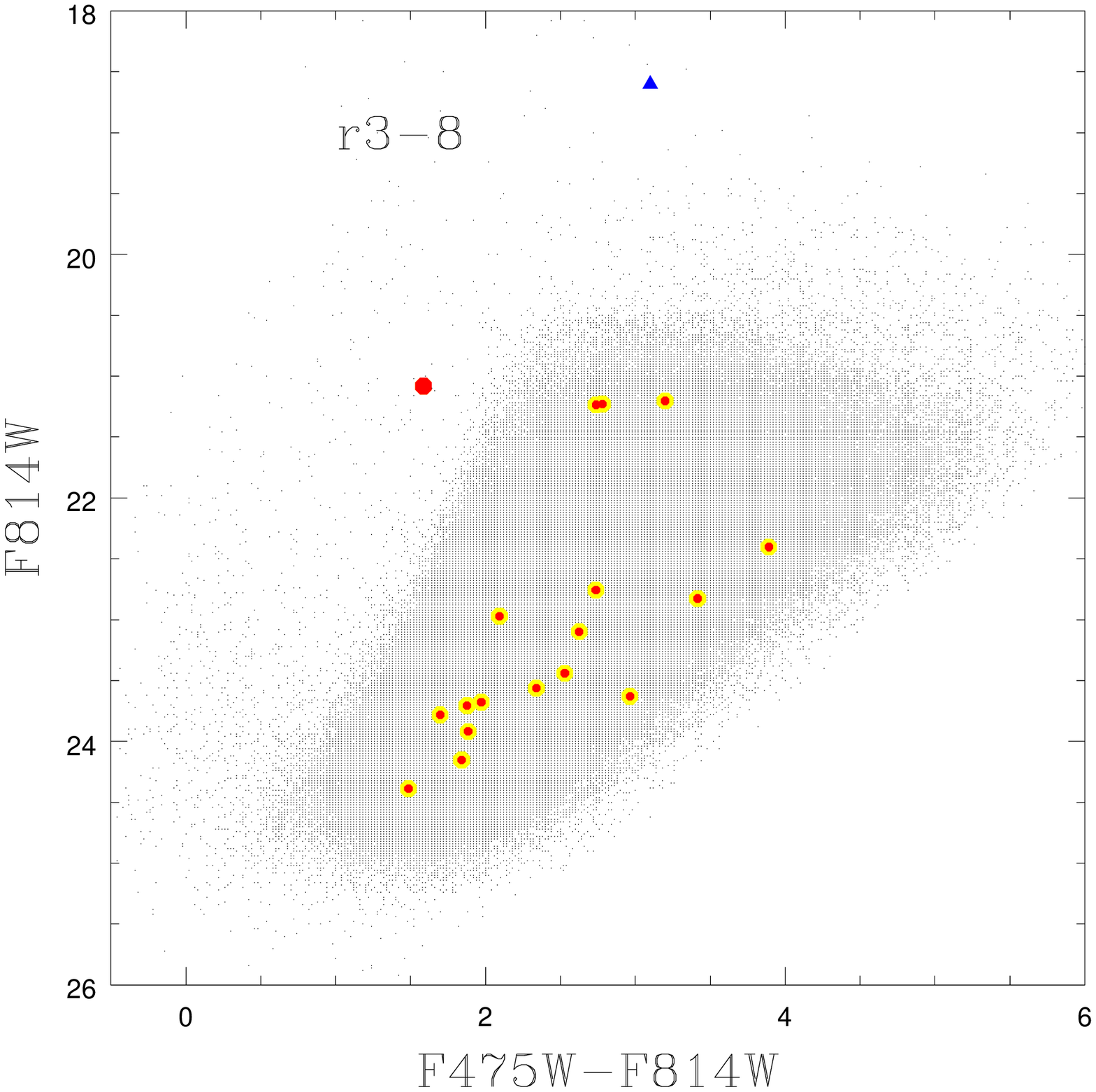}}\\
}
\hspace{0.2cm}
\parbox{7.5cm}{
\resizebox{7.5cm}{!}{\includegraphics{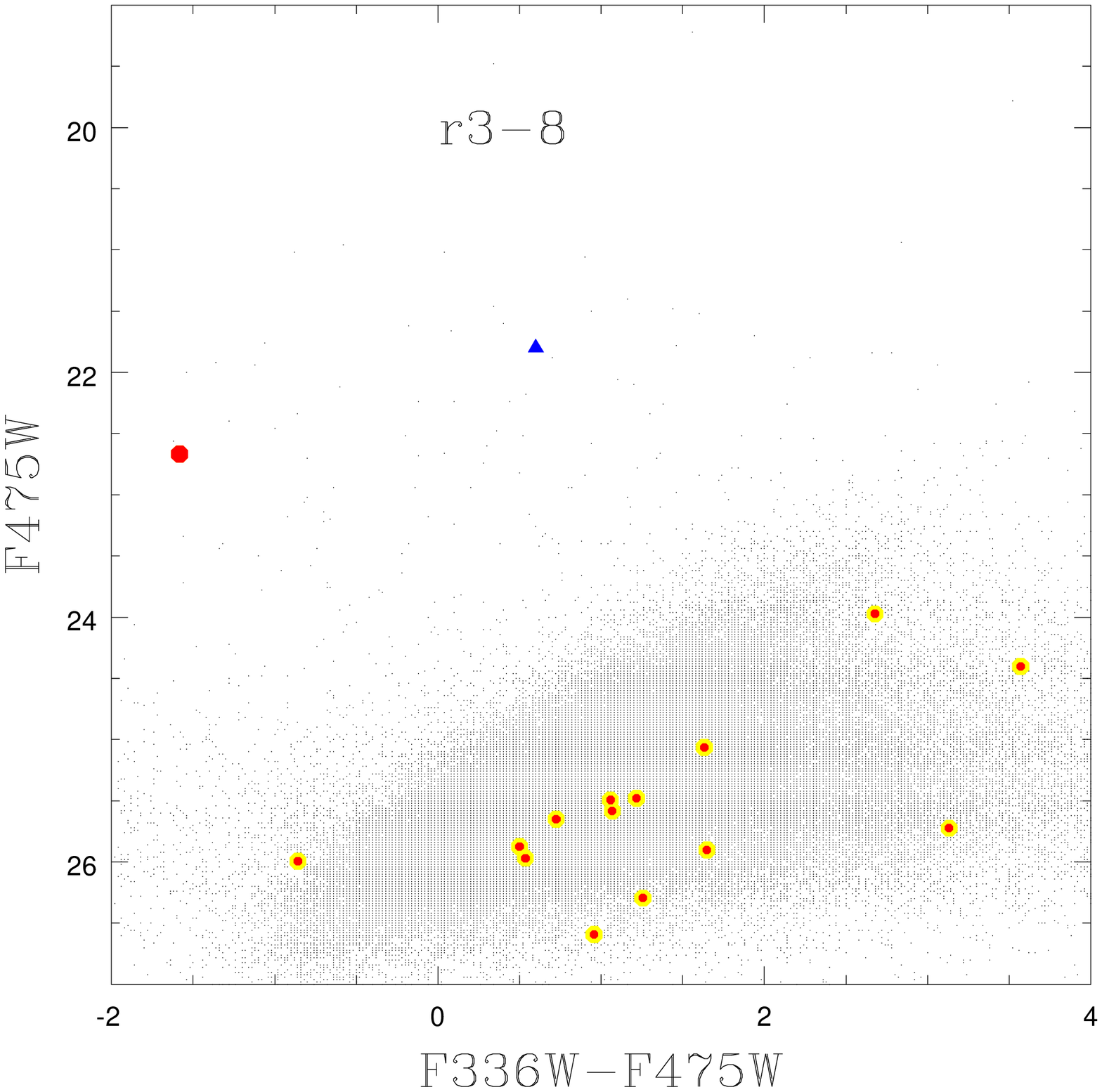}}\\
}
\caption{Position of the optical/U/UV 
 counterparts on the CMD of the fields of the persistent ``luminous''
 SSS,   based on the Dalcanton et al. photometry.
 The blue triangle represents the position of the symbiotic star SMC3
 as it would appear at the distance of M31. The bet candidates (``A'' for r2-12) are
 indicated by red dots. 
 } 
\label{cmd3.fig}
\end{figure*}
\begin{figure*}
   \centering
\parbox{6.7cm}{
\resizebox{6.7cm}{!}{\includegraphics{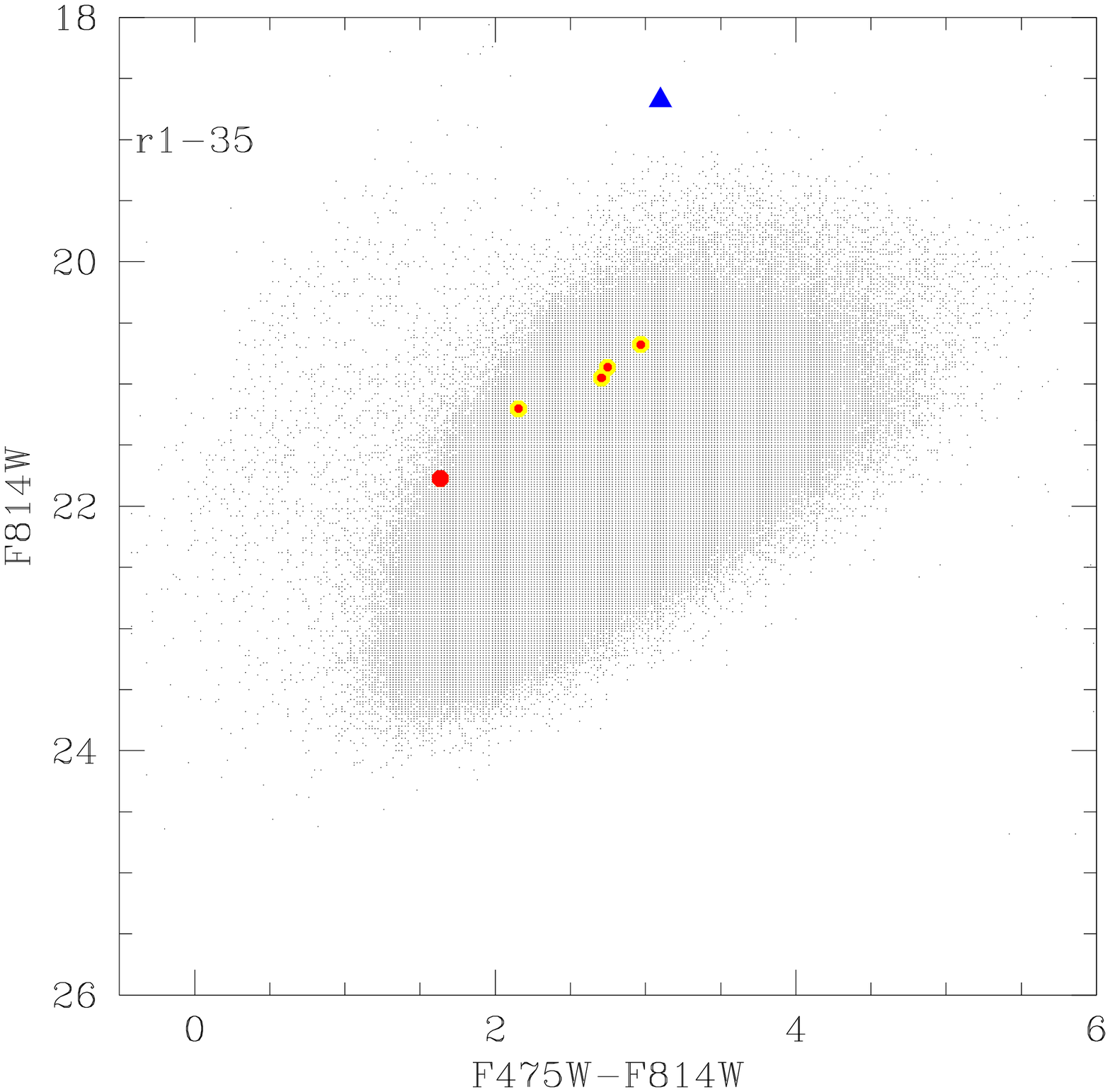}}\\
}
\hspace{0.1cm}
\parbox{6.7cm}{
\resizebox{6.7cm}{!}{\includegraphics{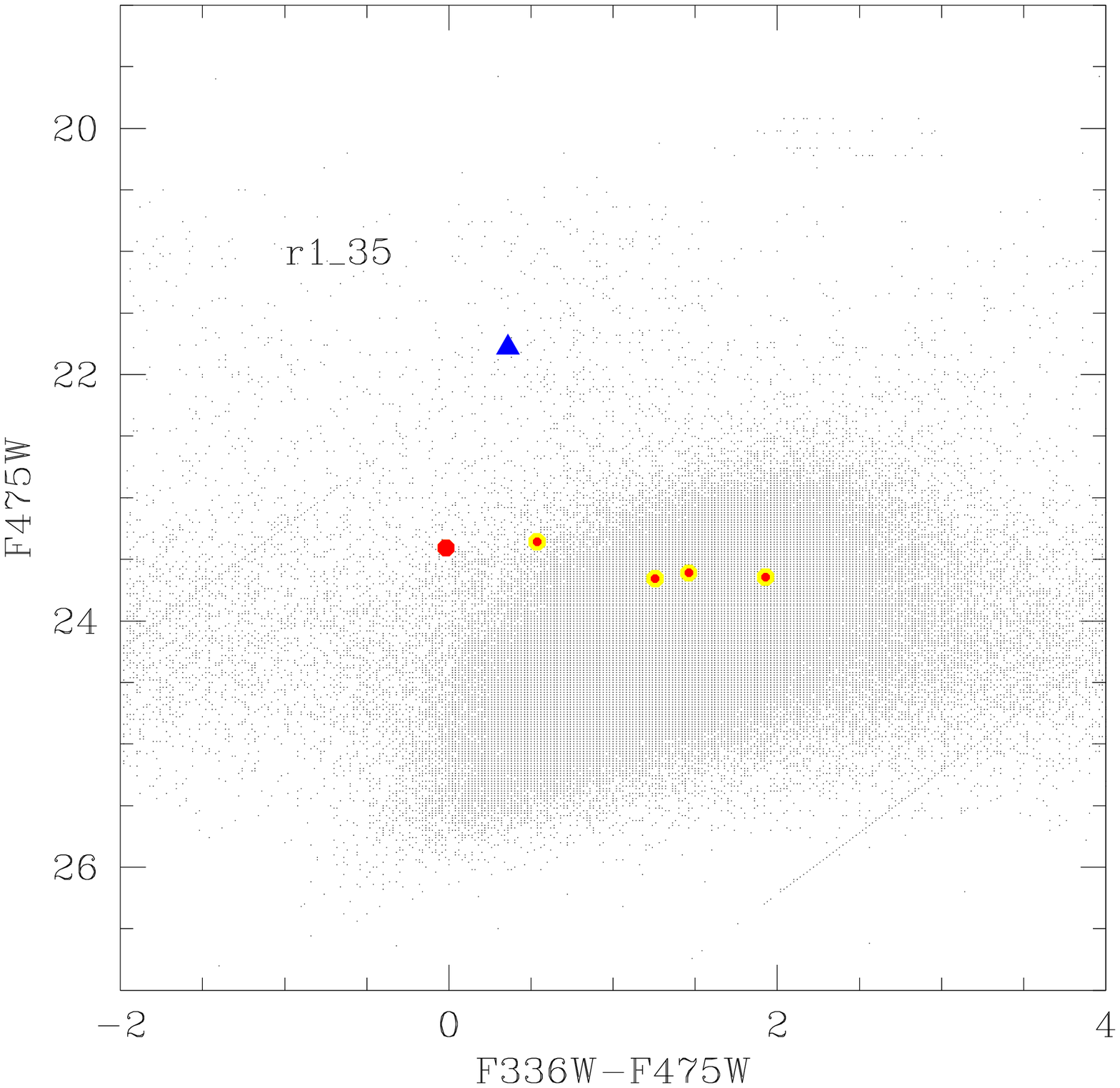}}\\
}
\parbox{6.7cm}{
\resizebox{6.7cm}{!}{\includegraphics{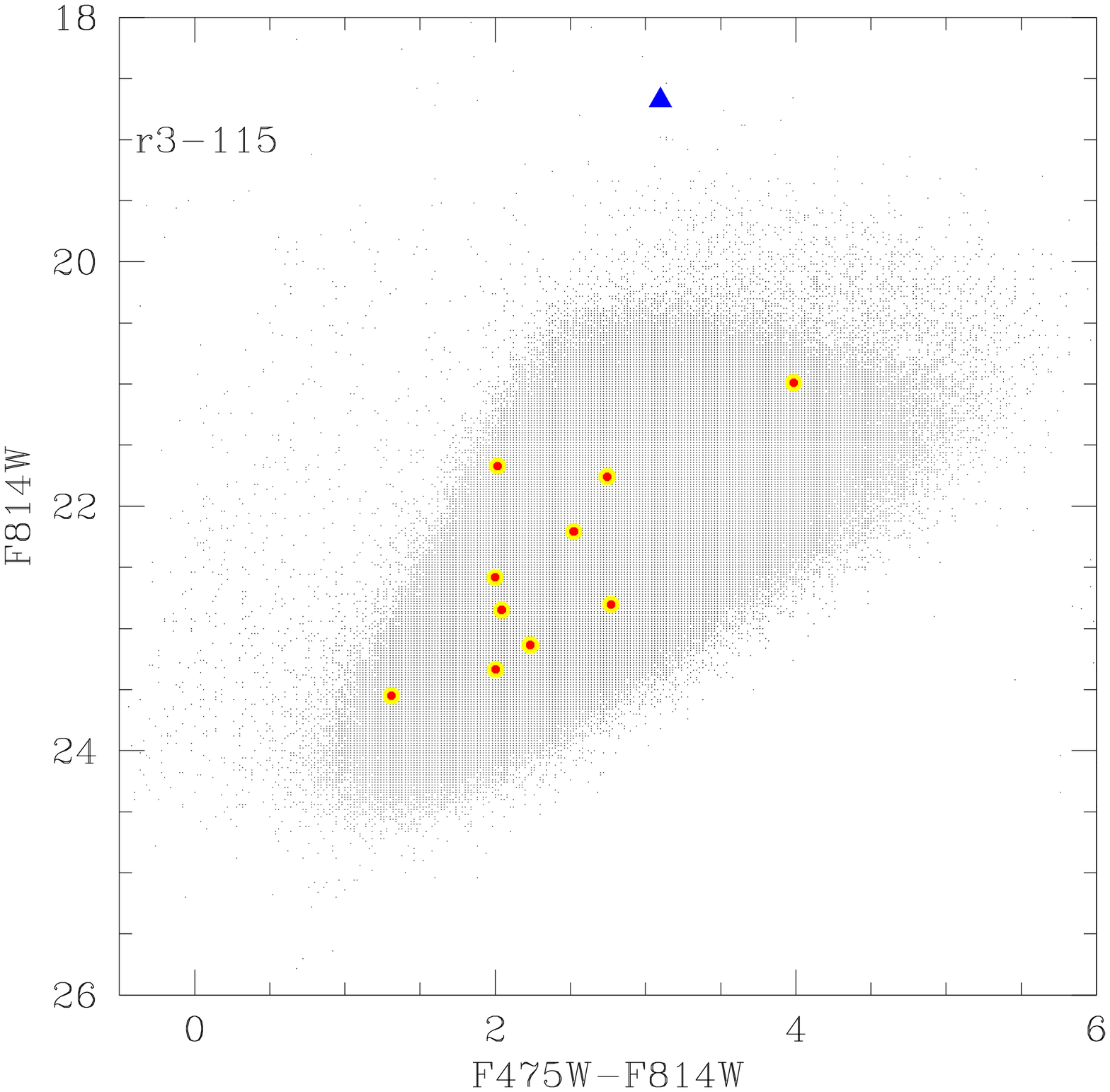}}\\
}
\hspace{0.1cm}
\parbox{6.7cm}{
\resizebox{6.7cm}{!}{\includegraphics{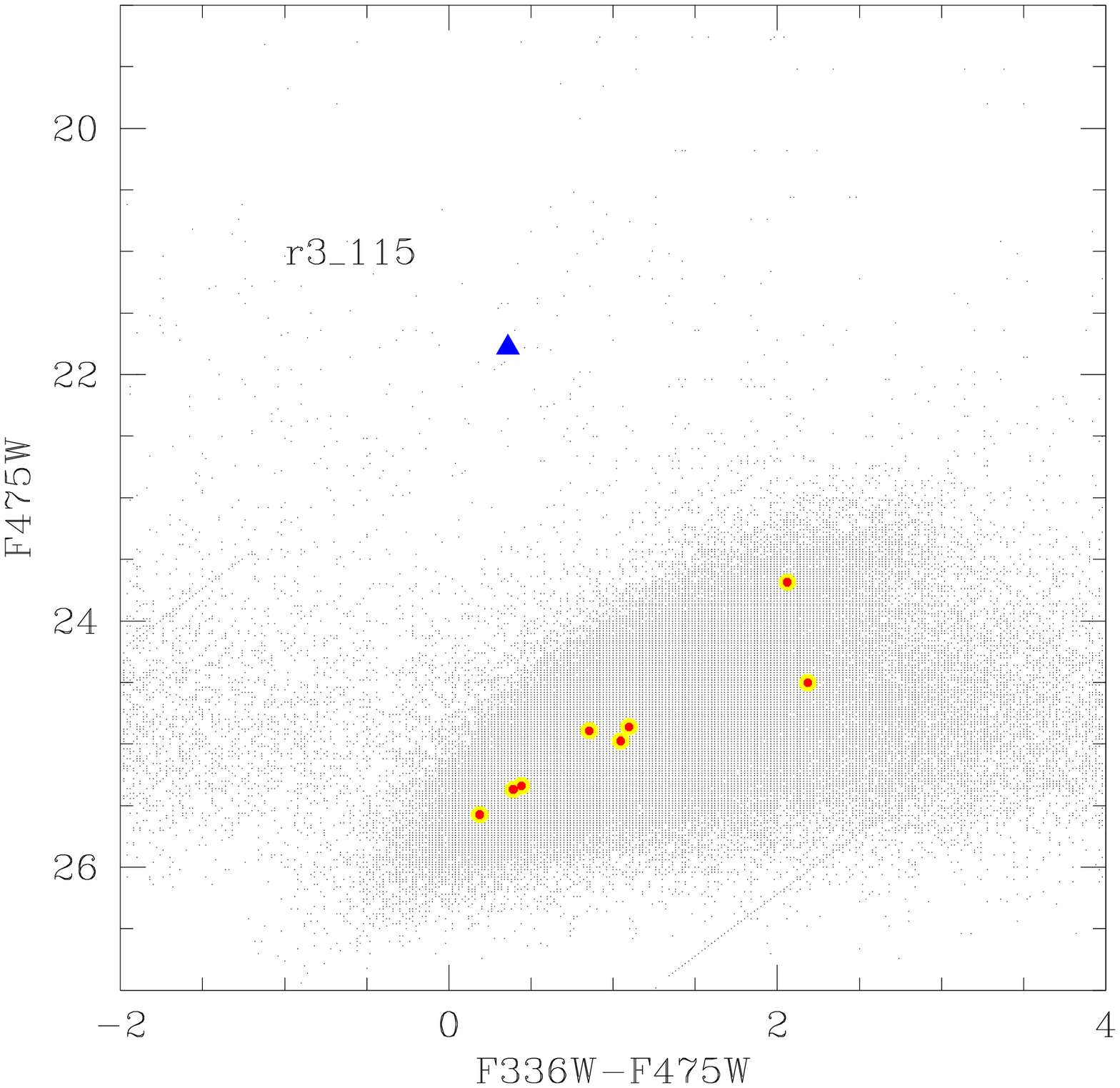}}\\
}
\parbox{6.7cm}{
\resizebox{6.7cm}{!}{\includegraphics{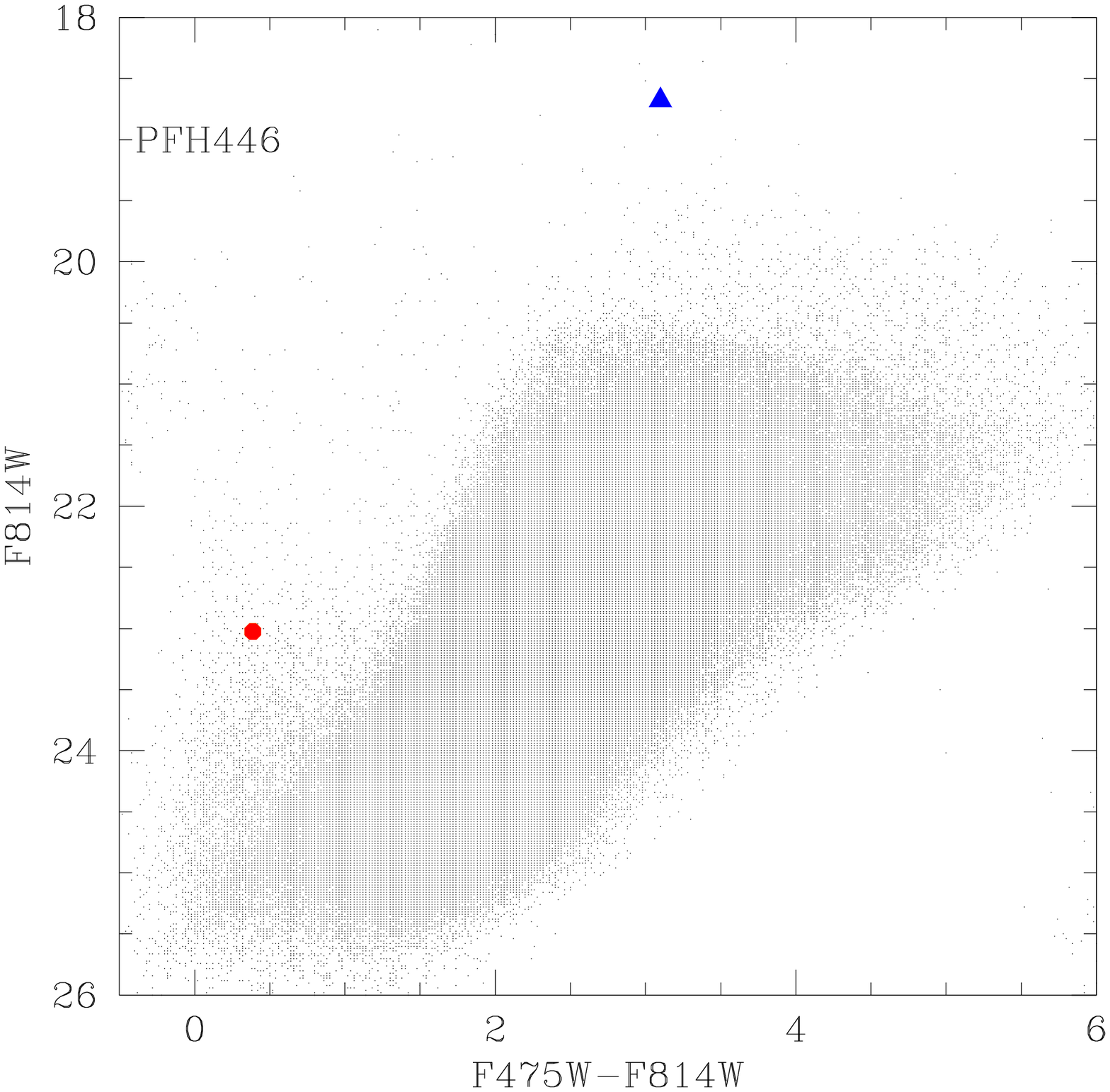}}\\
}
\hspace{0.1cm}
\parbox{6.7cm}{
\resizebox{6.7cm}{!}{\includegraphics{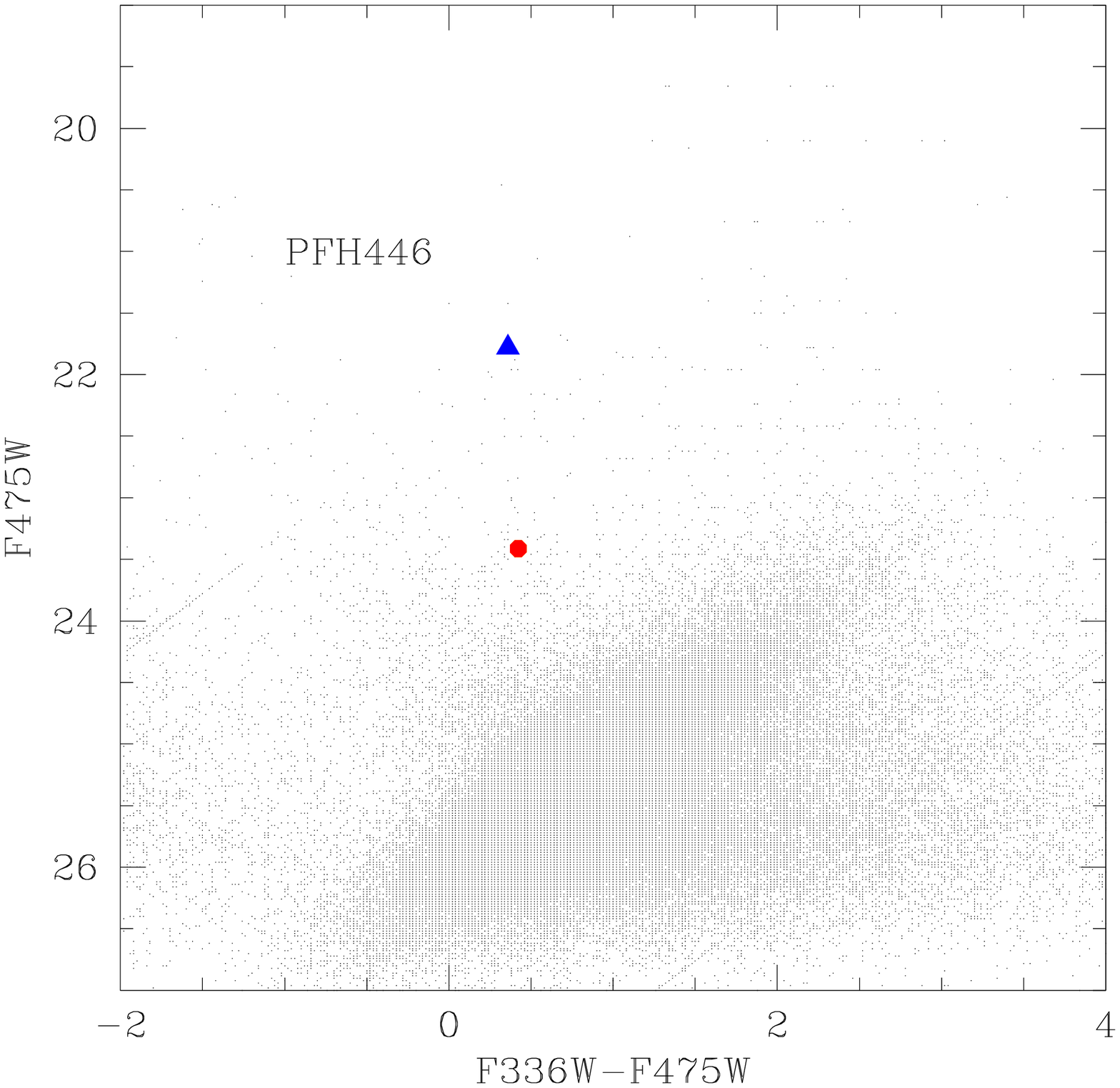}}\\
}
 \caption{Positions of the optical/U/UV candidates on the CMD
of the fields of the transient SSS, 
 based on the Dalcanton et al. photometry.
 The blue triangle represents the position of the symbiotic star SMC3 
as it would appear at the distance of M31,
 the red dots indicate the  candidate counterparts.
}
\label{cmd3.fig}
\end{figure*}

\section{Discussion and Conclusions}

The SSS in M31 offer a unique possibility to classify and study a statistically
 significant number of these X-ray sources at known distance. The SSS in binaries with
 characteristics similar to nova systems, but with steady instead
 of transient thermonuclear burning,
 may belong to three classes of single degenerate binaries evolving towards type Ia
 SN in single degenerates: 1) The \citet{vandenheuvel1992}) type binary, 
a massive WD with a more massive main sequence secondary, 2) Massive young
 binaries with a hydrogen burning WD, 3) Symbiotics.
The SSS we examined here are either in the core
 or are very close to the central region of M31,
 and not surprinsingly we have not found any optical object with
 the colors of high mass X-ray binaries or 
 B stars.  Our group of SSS 
 belongs to an old population, although in general  SSS as a class are clearly a mixture of young and
 old binaries.

 Objects with colors that may indicate  symbiotics,
are found in the error-circles of  r1-35 and PFH 446. There is a 
 more dubious ``candidate'' in the
 error circle of r2-54. PFH 446 is  in  a region 
which is not too crowded for future  spectroscopic follow-up, 
 but the proposed counterpart is fainter than 23rd magnitude. appears
 The ground based facilities of the future with
 adaptive optics  should be able to obtain optical spectra and
 prove the physical nature of these three objects. 

 Source r3-8 has a unique optical counterpart that is luminous
 in H$\alpha$ and is optically
 variable. It is a likely symbiotic because of
 the large IR luminosity, and even more luminous in the U band  than SMC3.
This SSS at times falls below detection limits for several
 weeks: this  may imply recurrent thermonuclear flashes 
 with or without mass ejection (see e.g. Fujimoto 1982). Recently, a recurrent nova has been
 observed in M31 with a recurrence period of only
 a year \citep{darnley2014, henze2014}, but thermonuclear flashes with a 
 shorter period  
 would occur without mass ejection. Another possibility is that the X-ray and optical
 variability are correlated and there is a periodic obscuration
 or eclipse during the orbital period, similar to the phenomenon observed in SMC 3.

   \begin{figure*}
   \centering
\parbox{7.5cm}{
\resizebox{7.5cm}{!}{\includegraphics{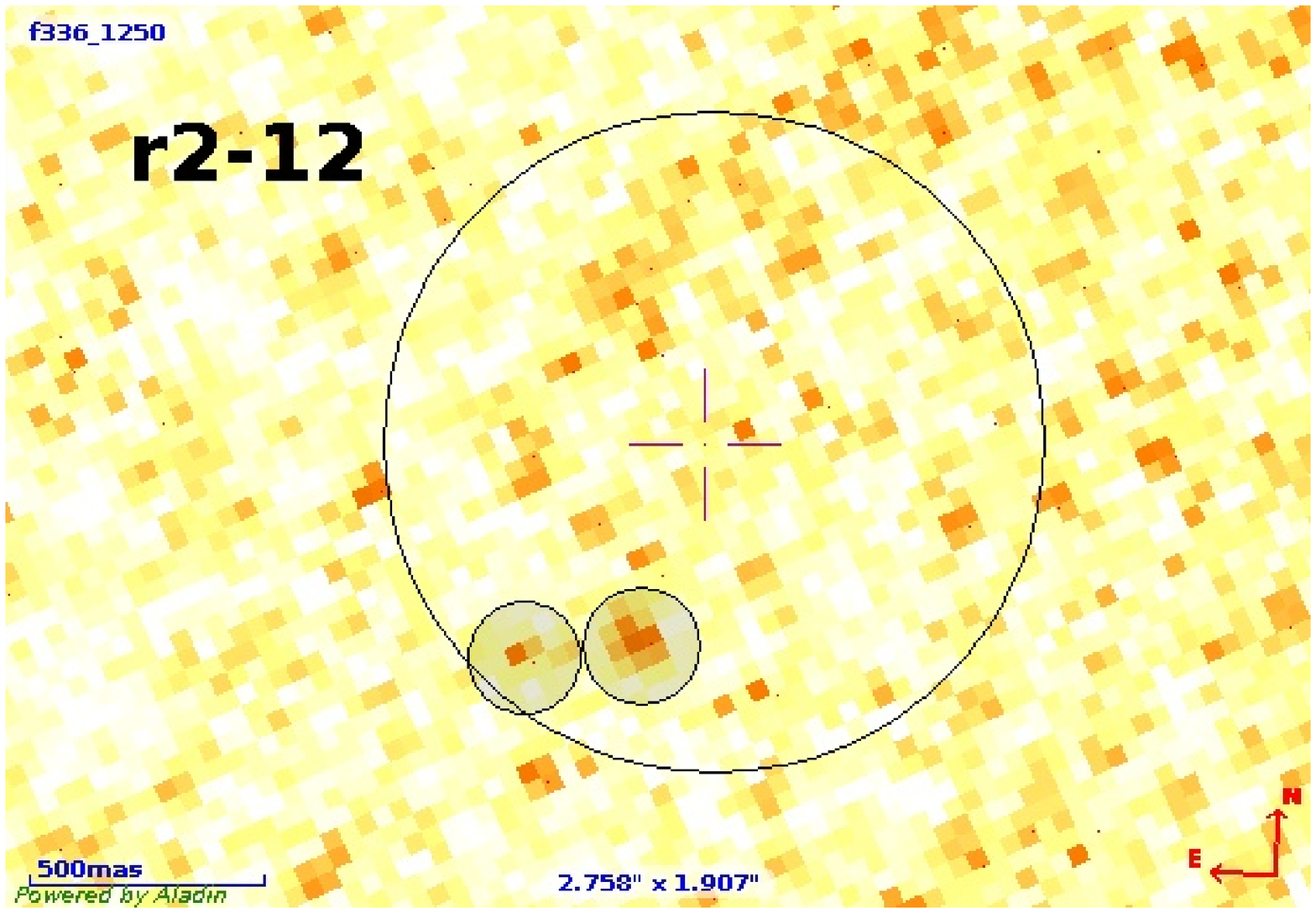}}\\
}
\hspace{0.2cm}
\parbox{7.5cm}{
\resizebox{7.5cm}{!}{\includegraphics{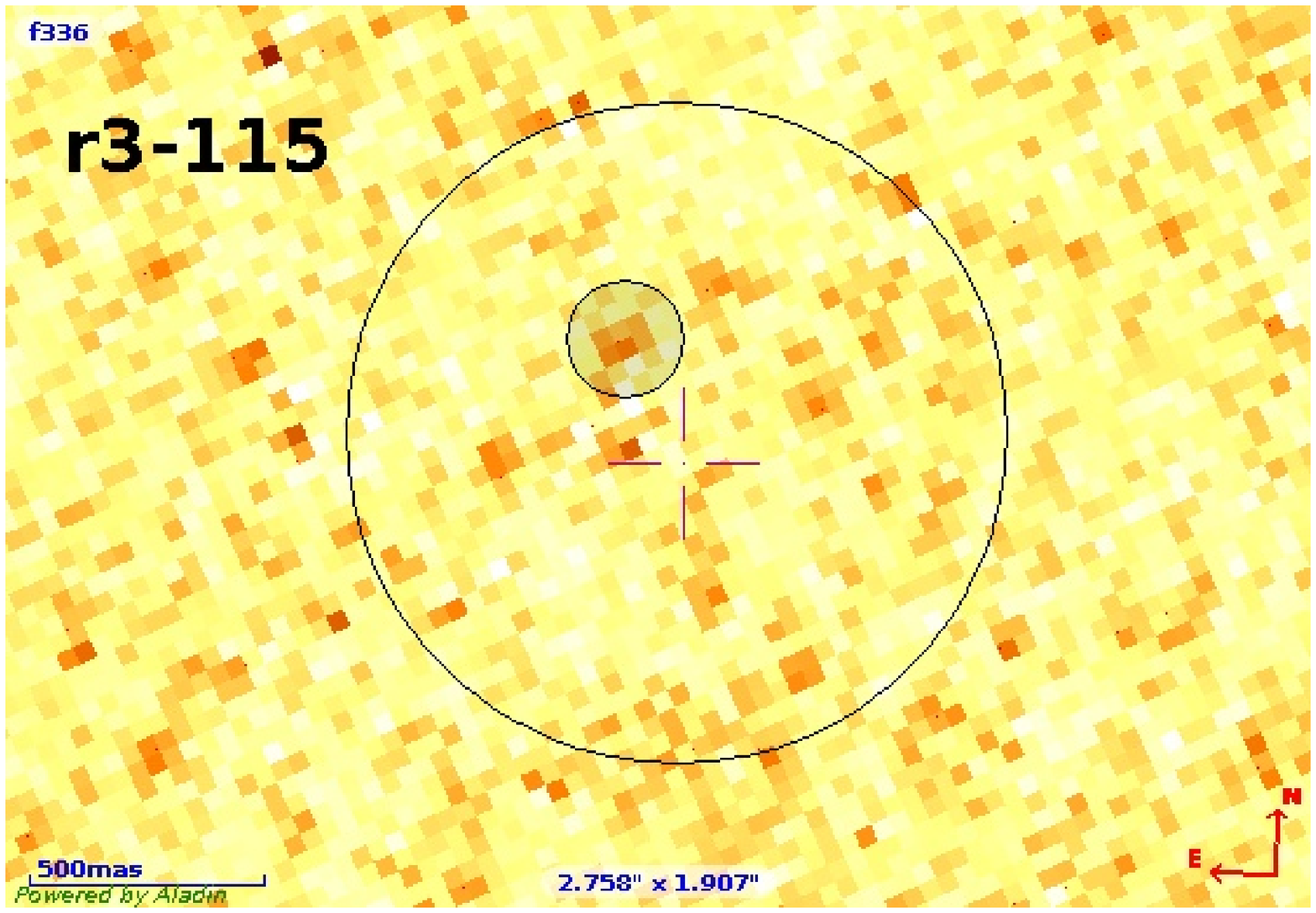}}\\
}
\parbox{7.5cm}{
\resizebox{7.5cm}{!}{\includegraphics{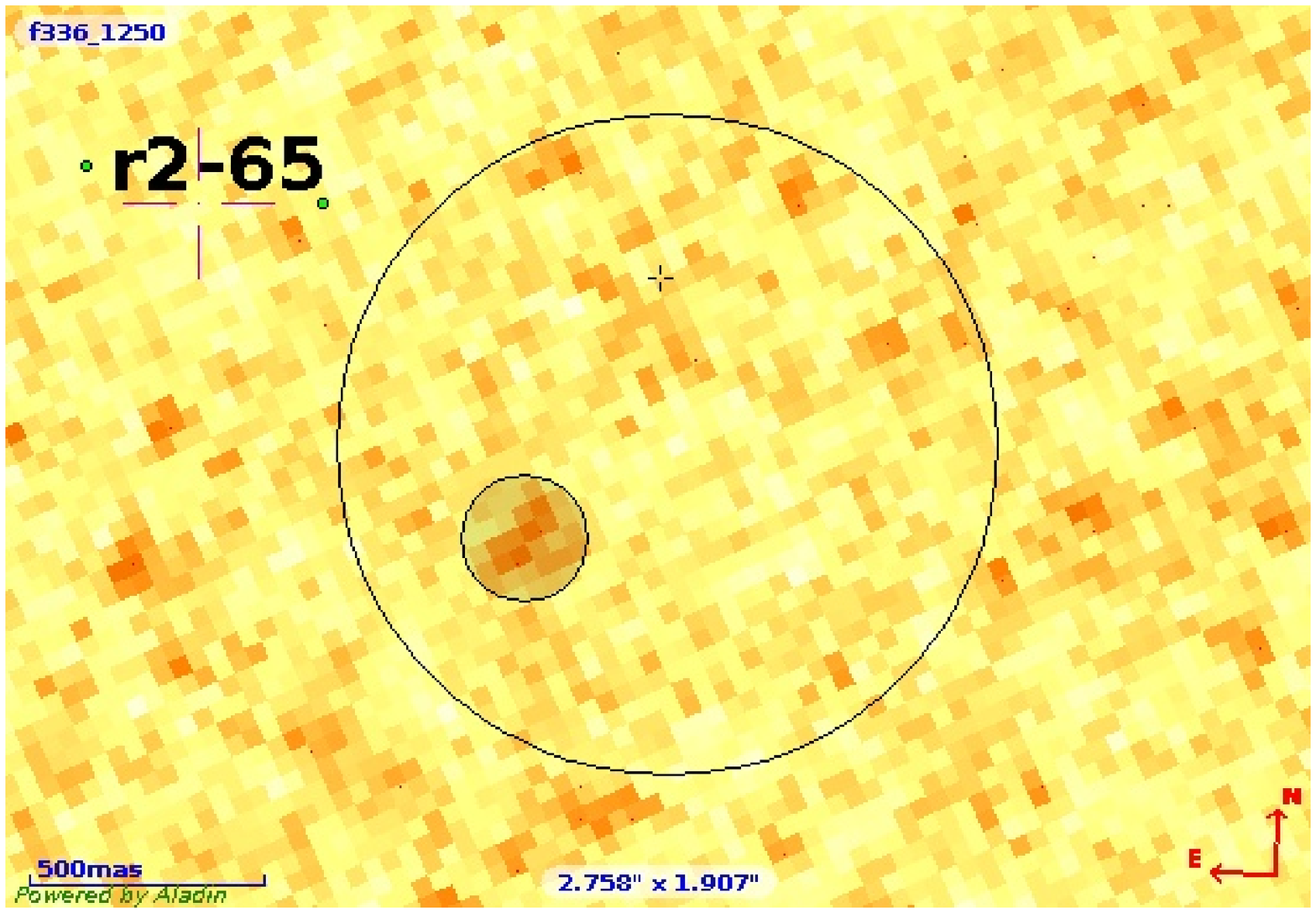}}\\
}
\hspace{0.2cm}
\parbox{7.5cm}{
\resizebox{7.5cm}{!}{\includegraphics{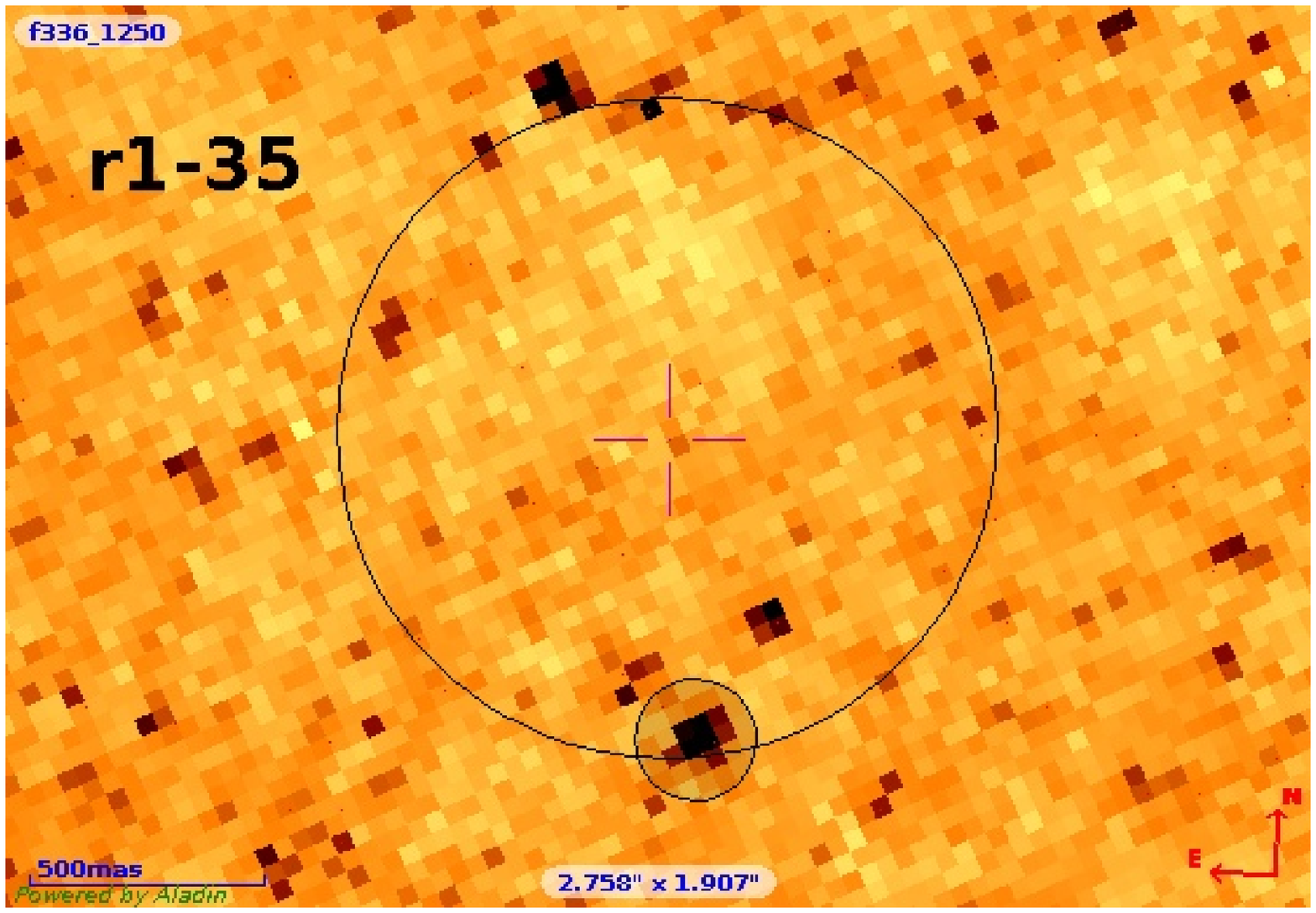}}\\
}
\parbox{7.5cm}{
\resizebox{7.5cm}{!}{\includegraphics{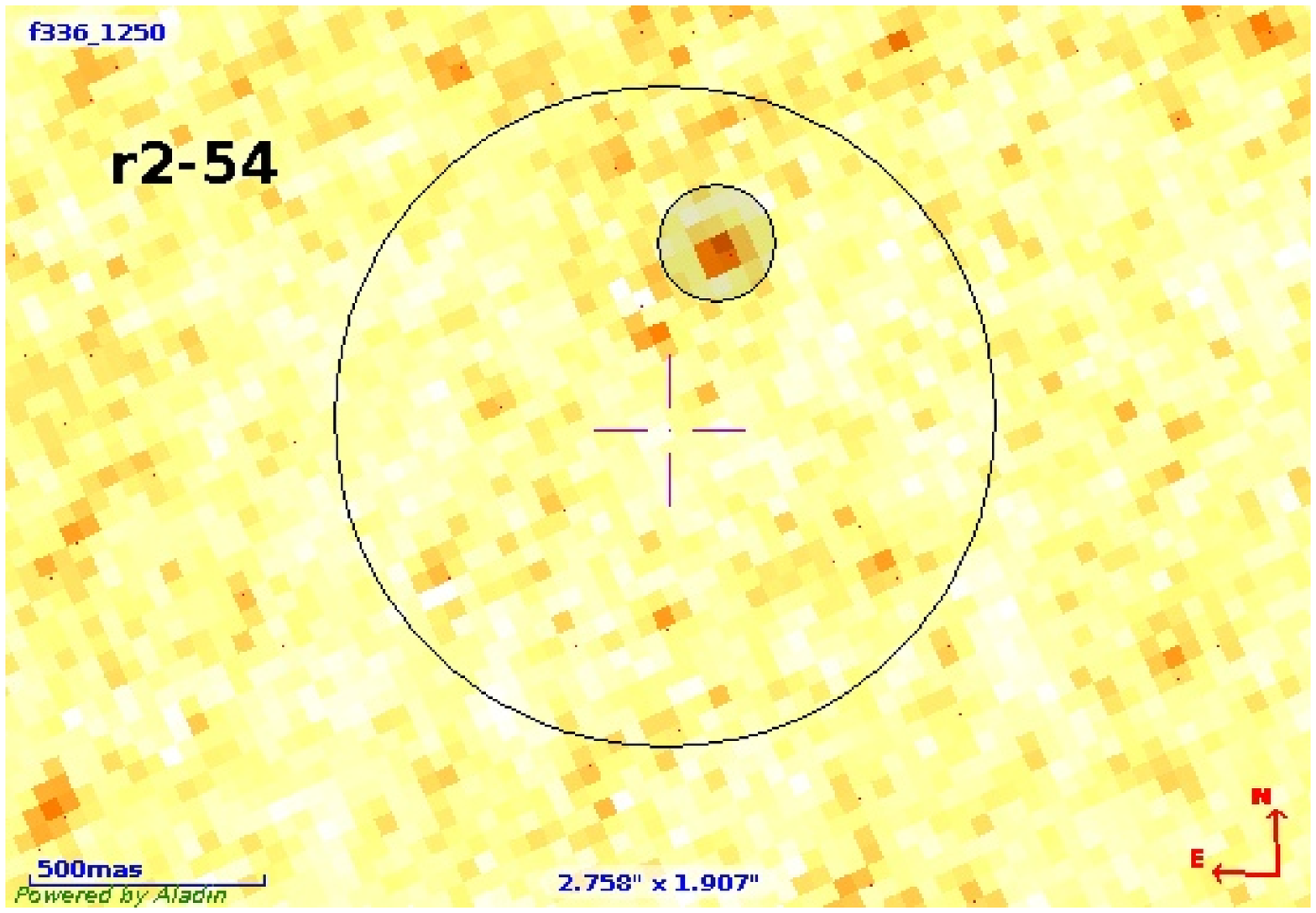}}\\
}
\hspace{0.2cm}
\parbox{7.5cm}{
\resizebox{7.5cm}{!}{\includegraphics{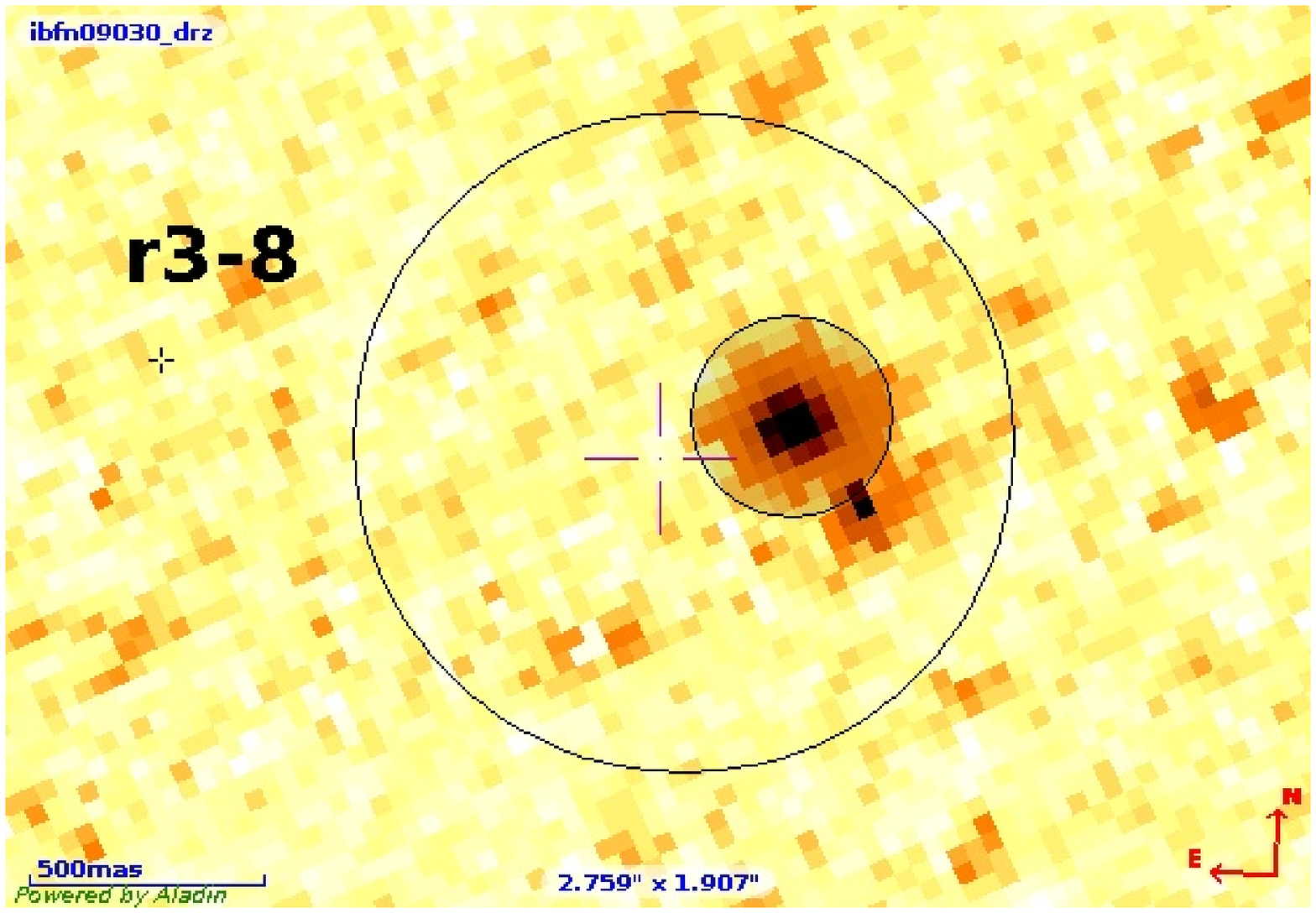}}\\
}
\parbox{7.5cm}{
\resizebox{7.5cm}{!}{\includegraphics{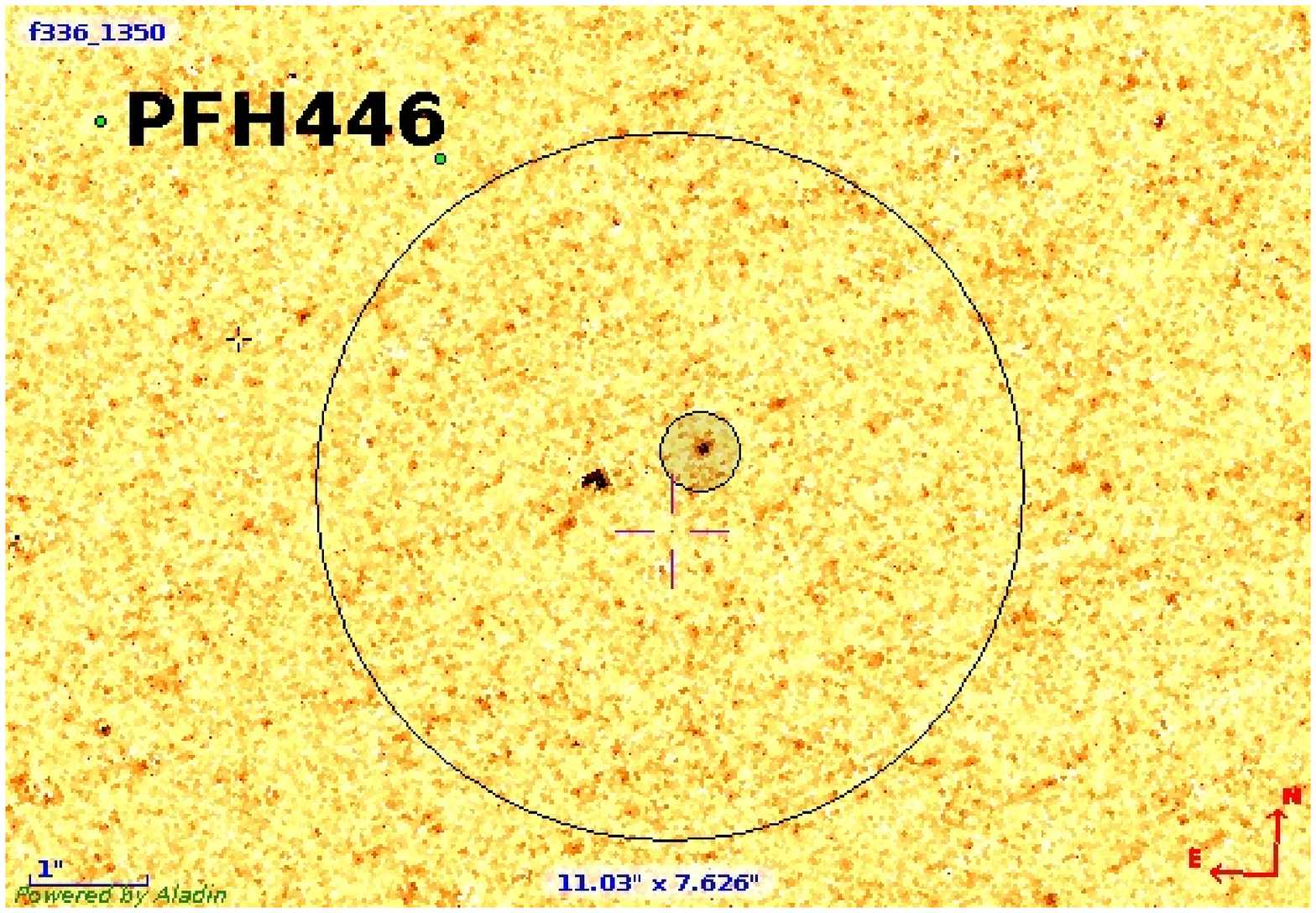}}\\
}
    \caption{
 The fields in the F336W filter images and 
 the spatial error circle of the X-ray positions of the SSS, with
 a 0.7'' radius for all the source except PFH 446, for which it is of 3''. 
 We re-registered the HST image of r3-8 with 2-MASS because there is an
additional systematic shift
 in the world coordinate system of the U/UV images of Brick 7 of the PHAT.
The small circles show the positions of the selected counterparts.
}
   \label{r3-8}
    \end{figure*}

Only ``regular'' red giants
 are in the error circles of r2-65 and r3-115.
 The low X-ray luminosity measured in the last years
 for r2-54 and r2-65  may rule out ongoing hydrogen burning,
 but it does not solve the mystery of their nature, because
 they are extremely luminous and soft. They may be SSS
 with  T$_{\rm eff} \leq$200,000 K, with  WD mass is between 0.4 and 0.7 M$_\odot$,
 effected  by copious intrinsic absorption 
  (see Starrfield 2012).
 The optical luminosity corresponds only to an upper limit 
of about 8 hours in in the orbital period diagram
 by van Teeseling et al., which would leave most short period CBSS undetected.
 Perhaps these two sources are quite
similar to SMC 13 \citep{Orio94, Crampton97}, but are more absorbed.

 The upper limits on the F435 W luminosity of
 the possible optical counterparts of r2-12 argue against 
 the scenario of a foreground, thermally emitting neutron star.
 We will also point out here that 
 a  $\simeq$ 3 minute pulsation period  is extremely unlikely for a neutron star. 
 Such long periods have only been observed in pulsars 
 with Be companions. While a thermally emitting
 cooling neutron star may also be a pulsar, Be companions  are very luminous
foreground counterparts  at optical
wavelengths \citep[][and references therein]{esposito2013} and would
 have not be missed. 

 The lack of a suitable optical/UV/H$\alpha$  counterpart for r2-12
 is an interesting puzzle, that
 can be understood in fact only if  this object is in a binary 
 orbital period below 169 minutes (2.8 hours), or else irradiation effects would make
 the disk sufficiently optically luminous to be detected in the PHAT.
This is the most peculiar
 of all the SSS we examined.  The detected periodic
 modulation is difficult to interpret without
 a WD, and it has high T$_{\rm eff}$ and luminosity.  
 The missing counterpart at other wavelengths leaves 
us with the intriguing possibility of a very short orbital period binary,
 like an AM CVn binary. 
We cannot rule out that the X-ray luminosity is
 due to helium instead of hydrogen burning (Bildsten et al. 2006).

 Another intriguing possibility is that the 
 217.7 s period is orbital in nature. The 
 shortest known period system is HM Cancri  (which is {\it not} an SSS)
and has a 321 s period. It is also known as
 RX J0806.3+1527 \citep{israel2002,roelofs2010}.
 The effective temperature of
 the WD in HM Cancri is about 140,000 K \citep{bildsten2006}, 
 much lower than the temperature inferred for the  r2-12 WD.
 In any case, r2-12 with its high effective
 temperature, high luminosity and a short orbital period,
is an ``extreme'' object
and it may represent an unusual, short-lasting state at the end of binary stellar evolution. 
By continuing the X-ray monitoring, we will test the physics of a very
 interesting source, probably a key object for studies of binary evolution
 and SN Ia progenitors' studies.
 
\bibliographystyle{mn2e}
\bibliography{biblio}
\begin{table*}
\centering
 \begin{minipage}{140mm}
\label{tab5.tab}
\caption{\small X-ray positions of the SSS, optical/UV HST images used to search for counterparts, and
 dates of X-ray detection or observation with no detection
 (either due to variability or to complete turn-off in X-rays. The H-number added for three sources is from the 
article by Hofmann et al. (2013). Uncertainties in the positions are discussed in Section 4.1}
\begin{tabular}{|l| l l |r | r r |r r| c|c|}
\hline
\multicolumn{1}{|l}{Object}&
\multicolumn{2}{|l}{Position}&
\multicolumn{1}{|l}{Field}&
\multicolumn{2}{|l}{Optical Datasets}&
\multicolumn{2}{|l}{UV Datasets}&
\multicolumn{1}{|c}{Detected}&
\multicolumn{1}{|c|}{Not Detected}\\
NAME &$ \alpha(J2000)$& $\delta(J2000)$ & (Brick 1)  & Filter & Exp(s) &
   Filter & Exp(s) & & \\
\hline

r2-54 & 00:42:38.77 & 41:15:26.44 & 11 &F814W&1700&  F336W  & 1250 & 2002 Oct  & often \\
H-106 &            &            &    &F475W&1890&  F275W  & 925  & 2005 Dec    & undetected \\ 
& & &                                &     &    &         &      &2006 Feb     & in shallower \\
& & &                                &     &    &         &      & 2010-2012   & images \\
\hline
r1-35& 00:42:43.11 & 41:16:04.2& 11&  F814W& 1700&  F336W &  1250&2001 Oct & 2004-2012 \\
H-146 & & &                             &  F475W& 1890&  F275W & 925  &  & \\
\hline
r2-65& 00:42:47.17 &41:14:13.2 & 16&  F814W&  1700&   F336W&   1250&2001 Oct & often \\
H-194 & & &                           &  F475W&  1890&   F275W&    925& 2008-2012  & undetected \\
\hline
r2-12 & 00:42:52.50 &41:15:40.1 & 10&  F814W&  1700& F336W&   1250 & 1979-2012 & \\
H-229 & & &                            &  F475W&  1890&  F275W&    925 & & \\
& & &                            & F658N &  2700 &       &        & & \\
\hline
r3-115& 00:43:06.98 & 41:18:10.5 & 03&  F814W&  1505&   F336W&   1350&  2001-2002 & 2004-2012  \\ 
H-273 & & &                           &  F475W&  1710&   F275W&   1010& &  \\
& & &         &       &      & &          & & \\
\hline
r3-8 &  00:43:18.92 & 41:20:16.7 & (B.7) & F475W &  1720 & F336W & 1300      & 1990-2012 & ``off'' states \\ 
H-294 &             &            & (B.7) &  F814W & 1520 & F275W & 1050      &           & (few weeks)  \\
\hline
PFH 446 & 00:43:25.56 & 41:16:17.1 &  01&  F814W&  1505&   F336W&   1350 & 2002 &2004-2012 \\
& & &                             &  F475W&  1710 & F275W&   1010 & &\\
\hline
\end{tabular}
\end{minipage}
\end{table*}

\normalsize
\begin{table*}
  \begin{center}
  \caption{Absolute, dereddened magnitudes and dereddened color indexes 
 for three known symbiotic SSS in the Local Group,  
 for the ``bluest'' object corresponding to the r2-12 {\sl Chandra}
 position, and of of the
 suggested candidate optical counterparts of r3-8, PFH 446, r1-35, and r2-54.
 The Johnson magnitudes are grouped together with the HST filter covering
 the (mostly) overlapping wavelength range. 
 From the 4th to the 8th column, the dereneddene color indexes are in the overlapping
 HST filters' bands, not in the Johnson bands.
The last row reports the assumed value for E(B-V).
 For dereddening we assumed Cardelli et al.'s extinction \citep{Card1988}.
 We did not add the
 smaller errors due to uncertainties in the distance modulus and the reddening. 
 The measurements of the already
 known Local Group SSS-symbiotics have very small errors, that are not reported here.
 We adopted a distance modulus 18.93$\pm$0.02 for the SMC \citep{Inno13},
 and 19.40$\pm$0.02 for Draco \citep{Bonanos08}.
}
  \label{tab1.tab}
 
\begin{tabular} {|ccccccccc|}
\hline 
Filter  & SMC3      & Lin 358   & Draco C1 & cand. r2-12  & cand. r3-8  & cand. PFH 446 & cand. r1-35&cand. r2-54\\
\hline
FW275   &           &          &          & +0.8$\pm$0.3       & -3.085$\pm$0.001       &  -0.9$\pm$0.1 & -1.03$\pm$0.08&0.3$\pm$0.3 \\
\hline 
F336W   &           &          &          & -0.10$\pm$0.25       & -3.765$\pm$0.001       & -1.12$\pm$0.03   & -1.46$\pm$0.0.03&-0.69$\pm$0.06 \\
U       & -2.60 (1) &          & -0.44(3) &              & -3.155       &         &     &  \\
\hline
F475W   &           &          &          & -0.16$\pm$0.02        & -2.08$\pm$0.01       & -1.41$\pm$0.01   & -1.34$\pm$0.01 &-0.10$\pm$0.03\\ 
B       & -3.21 (1) & -2.74(2) & -1.09(4) &              & -1.637       &         & &\\ 
\hline
V       &-4.46 (1)  &-3.40(2)  & -2.34(3) &              &              &  && \\
\hline
R       &           &           & -3.00(5)&              & -3.00        &    && \\
H$\alpha$ &         & -4.89(2)  & -3.18   &              & -2.710       &    &&   \\
\hline
F814W   &           &           &         & -1.06 $\pm$0.02       & -3.509$\pm$0.001       & -1.60$\pm$0.02 & -2.82$\pm$0.01&-3.168$\pm$0.005 \\
I       & -5.96(1)  &           &-3.79(3) &              & -3.599       &      & & \\
\hline
F110W   &           &           &         &           & -5.700$\pm$0.001  & -2.63$\pm$0.02  &   &-5.858$\pm$0.001  \\   
J       & -7.10(2)  &-6.53(2)   &-5.05(4) &           &         &        &      &  \\
\hline
(B-J)$_o$ & +3.89 & +3.79 & +3.83    & +7.15     & +3.618 & +0.91  & +4.29& \\
\hline
(B-V)$_o$ & +1.25 & +0.66 & +1.27     &              & -0.062  &  & &\\
\hline
(U-B)$_o$ & +0.41 &        & +0.62   & +0.06$\pm$0.27   & -1.68$\pm$0.01 & +0.28$\pm$0.04 & -0.18$\pm$0.04 &-0.59$\pm$0.09  \\
\hline
E(B-V)  & 0.08  &  0.08  & 0.03    & 0.08         & 0.08  & 0.10 & 0.08& 0.08 \\
\hline
\multicolumn{9}{|l|}{[1]\cite{boyer11},[2] 2MASS catalog, [3]\cite{kinemuchi08}, [4]\cite{adelman08},[5]\cite{cutri03}}\\
\hline
  \end{tabular}
 \end{center}

\end{table*}
\begin{table*}                                                                                                        \caption{$XMM-Newton$ observations' dates and duration and the measured periods.
The uncertainty is at the $1\sigma$ confidence level.}
\begin{center}
\begin{tabular}{ccccc}
\hline
\\
ObsID             & Start Date          & End Date                  & Duration           & Period \\
                  & yyyy-mm-dd hh:mm:ss & yyyy-mm-dd hh:mm:ss       &    s               &    s   \\
\hline                                                                                                            0112570401        & 2000-06-25 08:12:41 &  2000-06-25 20:59:33      &  46012  & $217.70\pm0.18$ \\
0112570101        & 2002-01-06 18:00:56 &  2002-01-07 11:52:53      &  64317  & $217.75\pm0.07$  \\
0202230201        & 2004-07-16 16:17:05 &  2004-07-16 21:54:04      &  20219  & $217.70\pm0.62$  \\
0405320501        & 2006-07-02 14:13:51 &  2006-07-02 20:19:04      &  21913  & $218.10\pm0.47$  \\
0505720301        & 2008-01-08 06:37:19 &  2008-01-08 14:10:58      &  27219  & $217.90\pm0.20$  \\
0551690601        & 2009-02-04 12:56:54 &  2009-02-04 20:25:31      &  26917  & $217.60\pm0.21$  \\
0600660301        & 2010-01-07 07:22:24 &  2010-01-07 12:11:00      &  17316  & $217.70\pm0.46$  \\
0650560301        & 2011-01-04 17:46:08 &  2011-01-05 03:03:03      &  33415  & $217.90\pm0.14$  \\
0674210601        & 2012-01-31 01:55:26 &  2012-01-31 09:08:03      &  25957  & $217.60\pm0.17$  \\
\hline
\end{tabular}
\label{tab:timing}
\end{center}
\end{table*}

\begin{table*} 
\caption{In this table, available in electronic format, we give the PHAT catalog 
 coordinates and magnitudes in the different filters of the objects 
in the error circles of 6 of the SSS. 
Column 1 and 2 are the position of the objects, column 3 is the offset
from the position of the sources listed in Table 1, columns
 4 and 5 are the magnitudes in the optical filters, column 6 and 7 are the magnitudes in the UV filters,
 and  column 8 is the infrared magnitude. The first raw shows the values for the ``blues'' candidate
 in each error circle, while the rest of the entries are in order of growing $\delta$.}
\begin{center}
\begin{tabular}{|l l l l l l l l|}
\hline
\multicolumn{8}{|l|}{objects in the r2-54 error circle}\\
\hline
 RA(J2000) &   DEC(J2000) &   offset(")& F475W  &    F814W &     F275W  &    F336W     &   F110W\\
           &              &           &  MAG   &     MAG  &      MAG     &   MAG     &     MAG\\
\hline
 00 42 38.77,& +41 15 26.88 &  0.37  &     24.851 &    21.422 &    25.251  & 24.160  &     18.682 \\   
 00 42 38.58,& +41 15 26.2  &  0.3   &     24.498 &    21.27  &    27.551  & 25.421  &     17.845 \\
 ... & ... &  ... & ...& ...& ...&...&...\\
\hline
\end{tabular}
\end{center}
\end{table*}

\label{lastpage}

\end{document}